\documentclass[twoside,12pt]{article}
\usepackage{epsfig}

\newcommand{\be}{\begin{equation}}
\newcommand{\ee}{\end{equation}}
\newcommand{\bea}{\begin{eqnarray}}
\newcommand{\eea}{\end{eqnarray}}

\newcommand{\gsim}{\raisebox{-0.7ex}{$\stackrel{\textstyle >}{\sim}$ }}
\newcommand{\lsim}{\raisebox{-0.7ex}{$\stackrel{\textstyle <}{\sim}$ }}
\def\si{^1 \hskip -0.03in S _0}
\def\siii{^3 \hskip -0.025in S _1}
\def\diii{^3 \hskip -0.03in D _1}

\begin{document}

\title{ \vspace{1cm} Nuclear Physics from Lattice QCD\footnote{Lecture
    presented at the Erice School on Nuclear Physics 2011: {\it From Quarks and
    Gluons to Hadrons and Nuclei}, organized by A. Faessler and J. Wambach.    } }
\author{Martin J. \ Savage\footnote{[NPLQCD Collaboration]} $^{1}$ \\
\\
$^1$Department of Physics,
  University of Washington, \\
Box 351560, 
Seattle, WA 98195, USA.
}
\maketitle
\begin{abstract} 
I review recent progress in the development of Lattice QCD into a
calculational tool for nuclear physics.
Lattice QCD is currently the only known way of ``solving'' QCD in the low-energy
regime, and it promises to provide a solid foundation for the
structure and interactions of nuclei directly from QCD.
\end{abstract}

\section{Introduction}
As discovered by the New Zealand physicist Ernest Rutherford,
a nucleus, labeled by its 
baryon number and electric charge, is at the heart of every atom.
Loosely speaking, nuclei are collections of protons and
neutrons that interact pairwise, with much smaller, but
significant, three-body interactions.
We are in the fortunate situation of knowing 
the underlying laws governing the strong interactions.  
It is the quantum
field theory called quantum chromodynamics (QCD),
constructed in terms of
quark and gluon fields with interaction determined by a local SU(3)
gauge-symmetry, along with quantum electrodynamics (QED),  that
underpins all of nuclear physics when the five relevant input
parameters, the scale of strong interactions $\Lambda_{\rm QCD}$, the three
light-quark masses $m_u$, $m_d$ and $m_s$, and the electromagnetic coupling
$\alpha_e$, are set to their values in nature.
It is remarkable that
the complexity of nuclei
emerges from ``simple'' gauge theories with just five input parameters.
Perhaps even more  remarkable is that  nuclei resemble
collections of nucleons and not collections of quarks and gluons.
By solving QCD, we will be able to predict, with arbitrary precision, nuclear
processes and the properties of multi-baryon systems (including, 
for instance,  the interior of
neutron stars).

The fine-tunings 
observed in the structure of nuclei and the interactions between nucleons
are peculiar  
and fascinating aspects of nuclear physics.
For the values of the input parameters that we have in our universe, 
the nucleon-nucleon (NN) interactions are fine-tuned 
to produce unnaturally large scattering lengths in both s-wave channels
(described by 
non-trivial fixed-points in the low-energy effective field theory (EFT)), 
and
the energy levels in the $^8$Be-system, $^{12}$C and $^{16}O$ are in ``just-so''
locations  to produce enough $^{12}$C to support life, and the subsequent
emergence and 
evolution of the human species.  
At a fundamental level it is important for us to determine the sensitivity of the
abundance of $^{12}$C to the light-quark masses and to ascertain the degree of
their fine-tuning.

Being able to solve QCD for the lightest nuclei,
using the numerical technique of Lattice QCD (LQCD), 
would allow for a partial unification of nuclear physics. 
It would be possible to ``match'' the traditional nuclear physics
techniques - the solution of the quantum many-body problem for neutrons and
protons using techniques such as 
No-Core Shell Model (NCSM), Greens function Monte Carlo
(GFMC), and others, to make predictions for the structure and interactions of
nuclei for larger systems than can be directly calculated with LQCD.
By placing these calculations on a fundamental footing, reliable predictions
with quantifiable uncertainties can then be made for larger systems.


\section{Lattice QCD Calculations of  Nuclear Correlation Functions}

Lattice QCD is a technique in which space-time is discretized into a
four-dimensional grid
and the QCD path integral over the quark and gluon fields at each
point in the grid is performed in Euclidean space-time
using Monte Carlo methods.
A LQCD calculation of a given quantity will deviate from its value in nature
because of the finite volume of the space-time (with $L^3\times T$ lattice points)
over which the fields exist, and
the finite separation between space-time points (the lattice spacing, $b$).
However, such deviations can be systematically removed by performing
calculations in multiple volumes with multiple lattice spacings, and
extrapolating using the
theoretically known functional dependences on each.
Supercomputers are needed for such calculations due to the number of space-time
points (sub grids  of which are distributed among the compute cores) 
and the Monte Carlo evaluation of the path integral over the dynamical fields.
In order for a controlled  continuum extrapolation, the lattice spacing  must be small enough
to resolve structures induced by the strong dynamics, 
encapsulated by $b\Lambda_\chi\ll 1$ where $\Lambda_\chi$ is the scale of
chiral symmetry breaking.
Further, in order to have the hadron masses, and also the scattering
observables, exponentially close to their infinite volume values, the lattice
volume must be large enough to contain the lightest strongly
interacting particle, encapsulated by $m_\pi L \gsim 2\pi$ where $m_\pi$ is the
mass of the pion and $L$ is the extent of the spatial dimension of the cubic
lattice volume (this, of course, can be generalized to  non-cubic volumes).
Effective field theory (EFT) descriptions of these observables exist
for $b\Lambda_\chi\lsim  1$ (the Symanzik action and its translation into
$\chi$PT and other frameworks) and $m_\pi L \gsim 2\pi$ (the p-regime of
$\chi$PT and other frameworks).
The low-energy constants in the appropriate EFT are fit to the results of the
LQCD calculations, which are then used to take the limit $b\rightarrow 0$ and
$L\rightarrow\infty$.
As the computational resources available today for LQCD calculations are not
sufficient to be able to perform calculations at the physical values of the
light quark masses in large enough volumes and at small enough lattice
spacings, 
realistic present day calculations are performed at light
quark masses that yield pion masses of $m_\pi\sim 200~{\rm MeV}$.
Therefore, present day calculations require the further extrapolation of
$m_q\rightarrow m_q^{\rm phys}$, but do not yet include strong isospin breaking or
electromagnetism.
In principle, the gluon field configurations that are generated 
in LQCD calculations
can be used to
calculate an enormous array of observables, spanning the range from particle to
nuclear physics.  In practice, this is becoming less common, largely due to the
different scales relevant to particle physics and to nuclear physics.
Calculations of quantities involving the pion with a mass of 
$m_\pi\sim 140~{\rm  MeV}$ are substantially different from those of, say, the
triton with a mass of $M(^3{\rm H})\sim 3~{\rm GeV}$, and with the typical scale of
nuclear excitations being $\Delta E \sim 1~{\rm MeV}$.
Present day dynamical LQCD calculations of nuclear physics quantities are performed with
$m_\pi\sim 400~{\rm MeV}$, lattice spacings of $b\sim 0.1~{\rm fm}$ and volumes
with spatial extent of $L\sim 4~{\rm fm}$.  Quenched calculations, which 
unfortunately cannot
be connected to nature, are typically performed in larger volumes as the
gauge-field configurations are less expensive to generate compared with dynamical configurations.

\begin{figure}[!ht]
\begin{center}
\begin{minipage}[t]{8 cm}
\epsfig{file=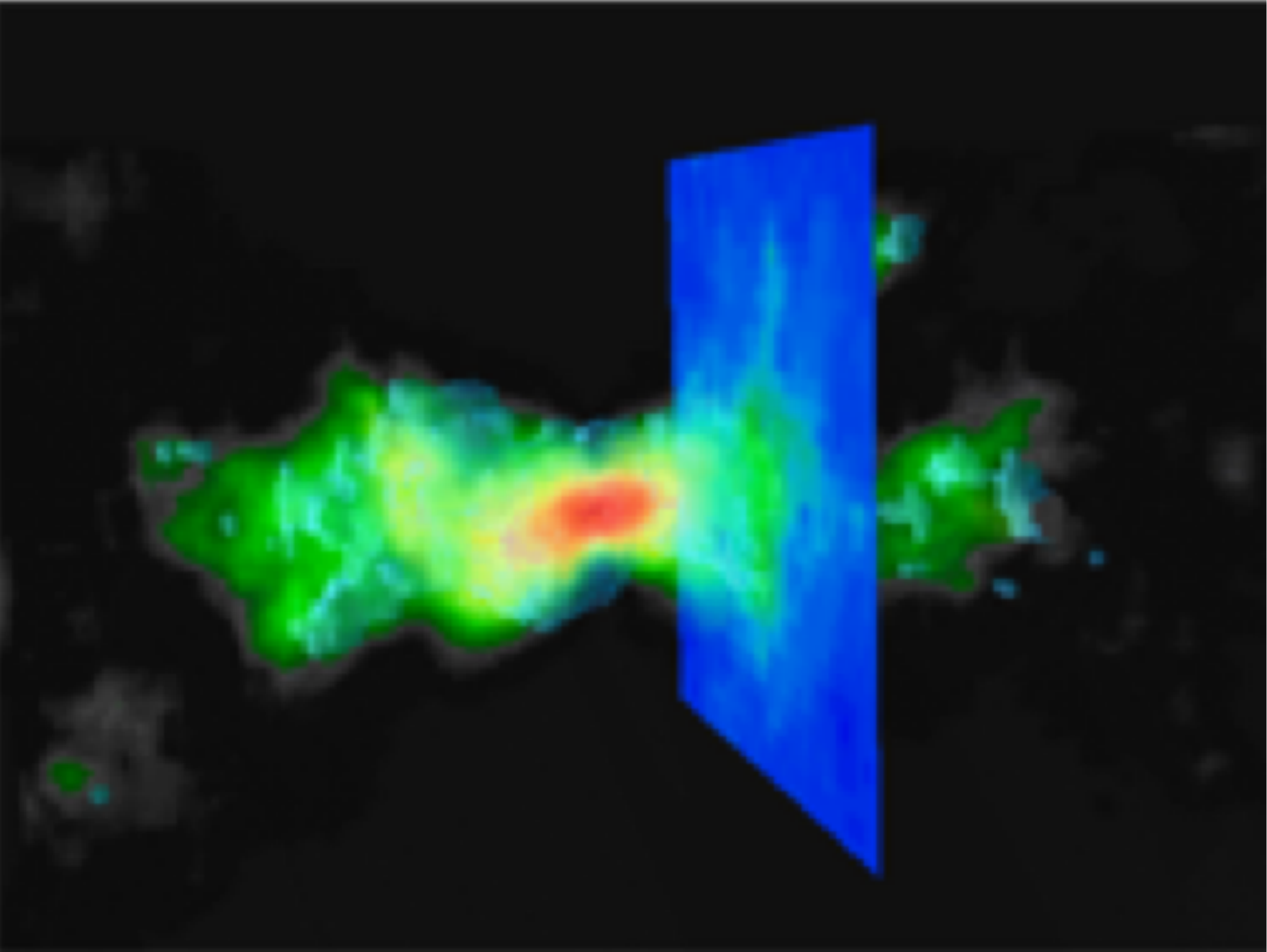,scale=0.5}
\end{minipage}
\begin{minipage}[t]{16.5 cm}
\caption{An example of (the real part of one component of)
a light-quark propagator.
The (blue) ``wall'' corresponds to the anti-periodic boundary conditions imposed in
the time direction.  [Image is reproduced with the permission of R. Gupta.]
\label{fig:prop}}
\end{minipage}
\end{center}
\end{figure}

LQCD calculations are approached in the same way that experimental efforts
use detectors to measure one or more quantities - the 
computer is equivalent to the accelerator
and the algorithms, software stack, and 
parameters of the LQCD calculation(s)
are the equivalent of  the detector.  The parameters, such as lattice spacing,
quark masses and volume, are selected based upon available computational
resources, and simulations of the precision of the calculation(s) required to
impact the physical quantity of interest, i.e. simulations  of the LQCD Monte
Carlo's are performed.  The size of the computational resources required for
cutting edge calculations are such that you only get ``one shot at it''.
A typical work-flow of  a LQCD calculation consists of three major
components.  The first component is the production of an ensemble of gauge-field
configurations which contain statistically independent samplings of the gluon
field configuration resulting from the LQCD action.  
The production of gauge-fields requires the largest partitions on the
leadership class computational facilities, 
typically requiring  $\gsim 128$K compute cores.
Present day calculations have
$n_f=0, 2, 2+1, 3, 2+1+1$ dynamical light quark flavors and use the Wilson, ${\cal
  O}(b)$-improved-Wilson, staggered (Kogut-Susskind), domain-wall or overlap
discretizations, each of which have their own ``features''.  
It is the
evaluation of the light-quark determinant 
(the determinant of a sparse matrix with dimensions
$\gsim 10^8\times 10^8$)
that consumes the largest fraction
of the resources.
Roughly speaking, $\gsim 10^4$ HMC trajectories are required to produce
an ensemble of $10^3$ decorrelated gauge-fields, but in many instances this is
an under-estimate.
For observables involving
quarks, a second component 
of production
is the determination of the light-quark
propagators on each of the configurations. 
The light-quark
propagator from a  given source point is determined
by an iterative inversion of the quark two-point function, 
using the conjugate-gradient (CG) algorithm or variants thereof such as BiCGSTAB,
an example of which is shown in Figure~\ref{fig:prop}.
During the last couple of years, the propagator production codes have been
ported to run on GPU machines in parallel.  The GPU's can perform propagator
calculations faster than standard CPU's by one or two orders of magnitude, and
have led to a major reduction in the statistical uncertainties in many calculations.
There have been numerous algorithm developments that have also reduced the
resources required for propagator production, such as the implementation of
deflation techniques and the use of multi-grid methods.
The third component of a LQCD calculation is the production of correlation
functions from the light-quark propagators.  This involves performing all of
the Wick contractions that contribute to a given quantity.  
The number of contractions required for computing a
single hadron correlation function is small.
However,  to acquire long plateaus
in the effective mass plots (EMP's) that
persist to short times, L\"uscher-Wolff type 
methods~\cite{Michael:1985ne,Luscher:1990ck} involve the
computation of a large number of correlation functions resulting from different
interpolating operators, and the number of contractions
can  become large.
In contrast, the naive number of contractions required for  a nucleus quickly
becomes astronomically large ($\sim 10^{1500}$ for uranium), but symmetries in
the contractions greatly reduces this number.  
For instance, there are 2880
naive contractions contributing to the $^3{\rm He}$ correlation function, but
only 93 are independent.
As a light-quark propagator can give rise to a pion correlation function
($\sim e^{-m_\pi t}$) and a nucleon correlation
function ($\sim e^{-m_N t}$) (and many other hadronic correlation functions), 
it is clear that the propagator
contains a hierarchy of mass scales, and that significant cancellations between
various components of the propagator occur in nucleon (and hence
nuclear) correlation functions.  
When taking a high power of the propagator, rounding
errors can accumulate, and 
in some cases it becomes necessary to perform calculations
using ``arbitrary precision'' libraries - for instance {\tt apprec}.
A further consequence of the hierarchy of mass scales is that there is an
asymptotic signal-to-noise problem in nuclear correlation functions.  
The ratio of the mean value of the correlation function to the variance of the
sample from which the mean is evaluated degrades exponentially at large times.
However, this is absent at short and intermediate times and
the exponential degradation of the signal-to-noise in the 
correlation functions can be avoided.

\section{Lattice QCD Calculations of Multi-Hadron Systems}
A driver for the development of LQCD technology is the
reproduction of 
the high-precision, experimentally determined, nucleon-nucleon phase-shift data
(verifying the LQCD technology) and the subsequent predictions of comparable
precision for hyperon-nucleon (YN), hyperon-hyperon (YY) interactions, along
with three-baryon (including nnn) and higher-body interactions.  
When completed, the precision of these latter observables will greatly exceed
what is possible experimentally, and therefore refine our ability to
calculate the properties exotic states of matter such as hyper-nuclei and the
interior of neutron stars, and to reduce the uncertainty in the fusion cross
section of light nuclei. 
Unfortunately, the formalism that is currently in place that
allows for the use of LQCD to extract information impacting phenomenology is somewhat limited. 
In fact, 
beyond the direct calculation of binding energies,
it is presently limited to using L\"uscher's method to determine the
scattering amplitude below inelastic thresholds in $2\rightarrow 2$ processes~\footnote{
While it has been suggested that one can extract nuclear ``potentials'' from
LQCD calculations, these energy-dependent and lattice scheme dependent
quantities contain no more information than the energy-eigenvalues (and hence
the scattering phase shift determined at the energy-eigenvalues via L\"uscher's method).
It is important to recognize the fact that these ``potentials'' produce
scheme-dependent values of the scattering amplitude at energies away from
the energy-eigenvalues.  As such, when used, for instance, in a nuclear
many-body calculation, they will produce results that are not predictions of QCD.
},
with straightforward extensions to coupled channels systems.

\subsection{\it Euclidean Space Correlation Functions}
\label{sec:ESCF}
\noindent
Most Euclidean space correlation functions computed in LQCD calculations 
(suitably Fourier transformed) are the sums of exponential functions.
The arguments of the exponentials 
are the product of Euclidean time with the eigenvalues
of the Hamiltonian associated with eigenstates in the finite-volume 
that couple to the hadronic sources and sinks.  
For a lattice that has infinite extent in the time-direction, the
correlation function at large times becomes a single exponential dictated by the
ground state energy and the overlap of the source and sink with the
ground state.  As an example, consider the pion two-point function,
$C_{\pi^+}(t)$, generated by a source (and sink) of the form
$\pi^+({\bf x},t)=\overline{u}({\bf x},t)\gamma_5 d({\bf x},t)$,
\begin{eqnarray}
C_{\pi^+}(t) & = & 
\sum_{\bf x}\ \langle 0|\ \pi^- ({\bf x},t)\ \pi^+ ({\bf 0},0)\ |0\rangle
\ \ \ ,
\label{eq:singlepioncorrelator}
\end{eqnarray}
where the sum over all lattice sites at each time-slice, $t$, projects onto the
 ${\bf p}={\bf 0}$ spatial momentum states.
The source $\pi^+({\bf x},t)$ not only produces single pion states, but also
all states with the quantum numbers of the  pion.
More generally, the source and sink are smeared over lattice sites in the
vicinity of $({\bf x},t)$ to increase the overlap onto the ground state and
lowest-lying excited states.
Translating the sink operator in time via $\pi^+({\bf x},t)=e^{\hat H t}
\pi^+({\bf x},0)e^{-\hat H t}$,
and inserting a complete set of states,
gives~\footnote{The absence of external electroweak fields that
  may exert forces on hadrons in the lattice volume is assumed.}
\begin{eqnarray}
C_{\pi^+}(t) & = & 
\sum_n\ {e^{-E_n t}\over 2 E_n}\  \sum_{\bf x}\ \langle 0|\ \pi^- ({\bf
  x},0) |n\rangle 
\langle n|\pi^+ ({\bf 0},0) |0\rangle
\ \rightarrow\ A_0\ {e^{-m_\pi t}\over 2 m_\pi}
\ \ \ .
\label{eq:singlepioncorrelatorASYMP}
\end{eqnarray}
At finite lattice spacing, the correlation functions for Wilson
fermions remain sums of exponential functions, but for particular
choices of parameters used in the domain-wall discretization, the
correlation functions exhibit additional sinusoidally modulated
exponential behavior at short-times with a period set by the lattice
spacing~\cite{Syritsyn:2007mp}.
It is straightforward to show that the lowest energy eigenvalue
extracted from the correlation function in
Eqs.~(\ref{eq:singlepioncorrelator}) and
(\ref{eq:singlepioncorrelatorASYMP}) corresponds to the mass of the
$\pi^+$ (and, more generally, the mass of the lightest hadronic state
that couples to the source and sink) in the finite volume.  The masses
of stable single particle states can be extracted from a Lattice QCD
calculation with high accuracy as long as the lattice spatial extent
is large compared to the pion Compton-wavelength~\footnote{Finite-volume
effects are exponentially suppressed~\cite{Luscher:1985dn} by factors
of $e^{-m_\pi L}$ in large volumes.}.

\subsection{\it Hadronic Interactions, the Maiani-Testa Theorem and L\"uscher's
  Method}
\noindent
Extracting hadronic interactions from Lattice QCD calculations is
more complicated than the determination of the spectrum of stable particles.
This is encapsulated in the
Maiani-Testa theorem~\cite{Maiani:1990ca}, which states that S-matrix elements
cannot be extracted from infinite-volume Euclidean-space Green functions except
at kinematic thresholds.
This could be problematic from the nuclear physics perspective, as a main
motivation for pursuing Lattice QCD is to compute nuclear reactions
involving multiple nucleons.
Of course, it is clear from the statement of this theorem how it can be evaded,
Euclidean-space correlation functions are calculated at finite volume to extract
S-matrix elements, the formulation of which was known for decades in the
context of non-relativistic quantum mechanics~\cite{Huang:1957im} and extended
to 
quantum field theory
by L\"uscher~\cite{Luscher:1986pf,Luscher:1990ux}.
The energy of two particles in a  finite volume depends
in a calculable way upon their elastic scattering amplitude and their masses
for energies below the inelastic threshold. 
As a concrete example consider $\pi^+\pi^+$ scattering.
A $\pi^+\pi^+$ correlation function 
in the $A_1^+$ representation of the cubic group~\cite{Mandula:ut} 
(that projects onto the continuum s-wave state amongst others) is
\begin{eqnarray}
C_{\pi^+\pi^+}(p, t) & = & 
\sum_{|{\bf p}|=p}\ 
\sum_{\bf x , y}
e^{i{\bf p}\cdot({\bf x}-{\bf y})} 
\langle \pi^-(t,{\bf x})\ \pi^-(t, {\bf y})\ \pi^+(0, {\bf 0})\ \pi^+(0, {\bf 0})
\rangle
\ \ \ . 
\label{pipi_correlator} 
\end{eqnarray}
In relatively large lattice volumes, the energy
difference between the interacting and non-interacting two-meson states
is a small fraction of the total energy, which is dominated by the
masses of the mesons.  
This energy difference can be extracted from
the ratio of correlation functions, $G_{\pi^+ \pi^+}(p, t)$,
where
\begin{eqnarray}
G_{\pi^+ \pi^+}(p, t) & \equiv &
\frac{C_{\pi^+\pi^+}(p, t)}{C_{\pi^+}(t) C_{\pi^+}(t)} 
\ \rightarrow \ {\cal B}_0\ e^{-\Delta E_0\ t} 
\  \ ,
\label{ratio_correlator} 
\end{eqnarray}
and where the arrow denotes the large-time behavior of $G_{\pi^+ \pi^+}$. 
For calculations performed with $p=0$,
the energy eigenvalue,  $E_n$, and its deviation from the sum of the rest
masses of the particle, $\Delta E_n$, are related to a
momentum magnitude $p_n$ by
\begin{eqnarray}
\Delta E_n \ & \equiv & 
E_n\ -\  2m_\pi\ =\ 
\ 2\sqrt{\ p_n^2\ +\ m_\pi^2\ } 
\ -\ 2m_\pi \ .
\label{eq:energieshift}
\end{eqnarray}
To obtain $k\cot\delta(k)$, where $\delta(k)$ is the phase shift, the
square of $p_n$ is extracted from the
energy shift and inserted
into~\cite{Huang:1957im,Luscher:1986pf,Luscher:1990ux,Hamber:1983vu}
\begin{eqnarray}
k\cot\delta(k) \ =\ {1\over \pi L}\ {\bf
  S}\left(\,\left(\frac{k L}{2\pi}\right)^2\,\right)
\ \ ,\ \ 
{\bf S}\left(\, x \, \right)\ \equiv \ \sum_{\bf j}^{ |{\bf j}|<\Lambda}
{1\over |{\bf j}|^2-x}\ -\  {4 \pi \Lambda}
\ \ ,
\label{eq:energies}
\end{eqnarray}
where $k=p_n$, and 
which is only valid below the inelastic threshold. 
The regulated three-dimensional sum~\cite{Beane:2003da}
extends over all triplets of integers ${\bf j}$ such that 
$|{\bf j}| < \Lambda$ and the
limit $\Lambda\rightarrow\infty$ is implicit.
Therefore, by calculating the energy-shift, $\Delta E_n$,  of the two 
particles in the finite lattice
volume, the scattering phase-shift is determined at $\Delta E_n$.
In the absence of interactions between the particles,
the energy eigenstates in the finite volume  
occur at momenta ${\bf p} =2\pi{\bf j}/L$.  
Perhaps most important for nuclear physics is that this expression is valid
for large and  even infinite scattering lengths~\cite{Beane:2003da}.  
The only restriction is that the lattice volume be much larger than the
range of the interaction between the hadrons, which for two nucleons, is set by
the mass of the pion.

\subsection{\it Meson-Meson Scattering}

\noindent 
The low-energy scattering of pions and kaons, the
pseudo-Goldstone bosons of spontaneous chiral symmetry breaking,
provides a perfect testing ground for Lattice QCD calculations of
scattering parameters. There is little or no signal-to-noise problem 
in such calculations and
therefore highly accurate Lattice QCD calculations 
of stretched-isospin states
can be performed
with modest computational resources. Moreover, the EFTs which describe the
low-energy interactions of pions and kaons, including lattice-spacing
and finite-volume effects, have been developed to non-trivial orders
in the chiral expansion. 
\begin{figure}[!ht]
\begin{center}
\begin{minipage}[!ht]{8 cm}
\centerline{\epsfig{file=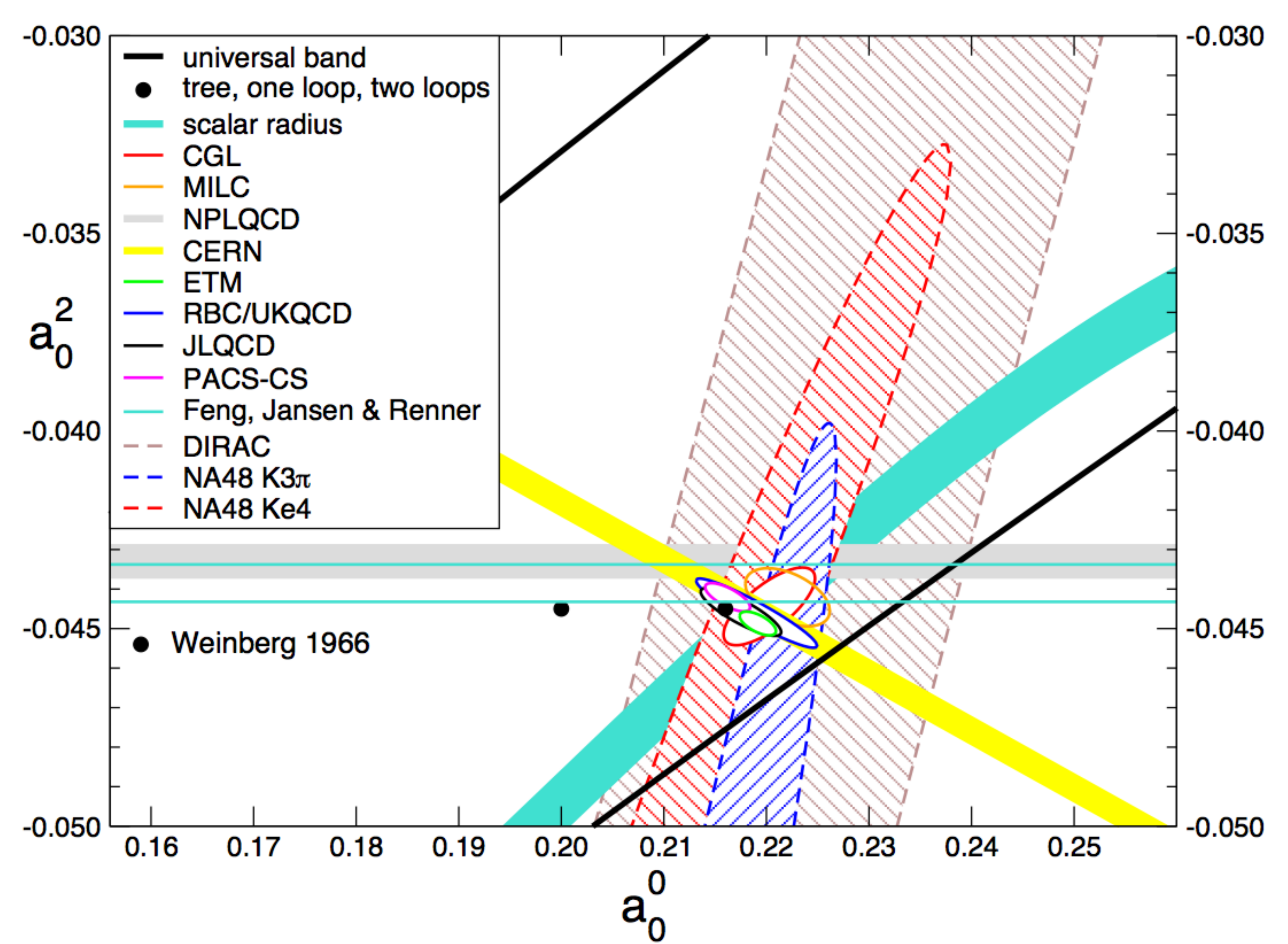,scale=0.25}\hspace{0.4cm}\epsfig{file=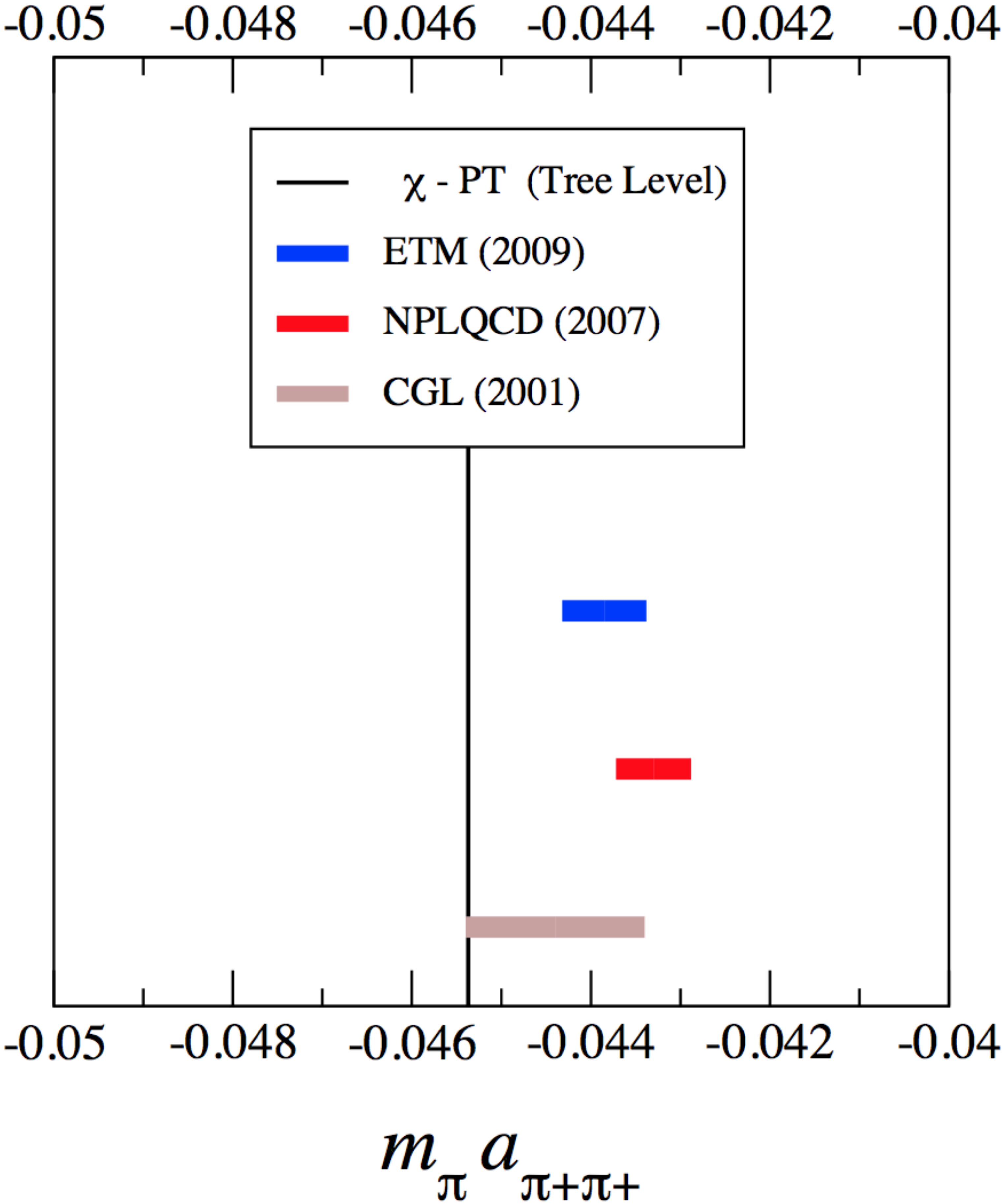,scale=0.195}}
\end{minipage}
\begin{minipage}[t]{16.5 cm}
\caption{
Present constraints on threshold s-wave $\pi\pi$ scattering. 
Noteworthy in the left panel ~\protect\cite{Leutwyler:2008fi} are
the red ellipse from the Roy equation analysis and 
the grey band from the direct Lattice QCD calculation of
the $\pi^+\pi^+$ scattering length, as discussed in the text.
The right panel shows the $\pi^+\pi^+$ scattering length results only.
[Image in the left panel is reproduced with the permission of H. Leutwyler.]
\label{fig:pipi2}}
\end{minipage}
\end{center}
\end{figure}
The $I=2$ pion-pion ($\pi^+\pi^+$) scattering length serves as a
benchmark calculation with an accuracy that can only be aspired to
in other systems.   The scattering
lengths for $\pi\pi$ scattering in the s-wave are uniquely predicted
at LO in $\chi$-PT~\cite{Weinberg:1966kf}:
\begin{eqnarray}
m_{\pi^+} a_{\pi\pi}^{I=0} \ = \ 0.1588 \ \ , 
\ \ m_{\pi^+} a_{\pi\pi}^{I=2} \ = \
-0.04537 
\ \ \ .
\label{eq:CA}
\end{eqnarray}
While experiments 
do not directly provide stringent
constraints on the scattering lengths, a determination of s-wave
$\pi\pi$ scattering lengths using the Roy equations has reached a
remarkable level of
precision~\cite{Colangelo:2001df,Leutwyler:2008fi}:
\begin{eqnarray}
m_{\pi^+} a_{\pi\pi}^{I=0} \ = \ 0.220\pm 0.005 \ \ , 
\ \ m_{\pi^+} a_{\pi\pi}^{I=2} \ = \ -0.0444\pm 0.0010
\ \ \ .
\label{eq:roy}
\end{eqnarray}
The Roy equations~\cite{Roy:1971tc} use dispersion theory to relate
scattering data at high energies to the scattering amplitude near
threshold. At present, Lattice QCD can compute $\pi\pi$ scattering
only in the $I=2$ channel with precision 
as the $I=0$ channel contains disconnected
diagrams which require large computational resources. 
It is of great interest to compare the precise Roy
equation predictions with Lattice QCD
calculations,
and  Figure~\ref{fig:pipi2} summarizes theoretical and
experimental constraints on the s-wave $\pi\pi$ scattering
lengths~\cite{Leutwyler:2008fi}. 
This is clearly a strong-interaction
process for which  theory has somewhat out-paced the challenging experimental
measurements.

Mixed-action $n_f=2+1$ Lattice QCD calculations, 
employing domain-wall valence quarks on a rooted staggered sea and 
combined with mixed-action $\chi$PT,
have predicted~\cite{Beane:2007xs}
\begin{eqnarray}
m_{\pi^+} a_{\pi\pi}^{I=2} & = &  -0.04330 \pm 0.00042
\ \ \ ,
\label{eq:nplqcd2}
\end{eqnarray}
at the physical pion mass,
where the statistical and systematic uncertainties have been combined
in quadrature.
The agreement between this result and the Roy equation
determination is a striking confirmation of the lattice methodology,
and a powerful demonstration of the constraining power of chiral
symmetry in the meson sector. 
However, lattice calculations at one or more smaller lattice spacings, 
and with different discretizations,
are
required to verify and further refine this calculation.
The ETM collaboration has  performed a $n_f=2$ calculation of
the $I=2$ $\pi\pi$ scattering length~\cite{Feng:2009ij}, producing a 
result extrapolated to the physical pion mass of
\begin{eqnarray}
m_{\pi^+} a_{\pi\pi}^{I=2} & = &  -0.04385 \pm 0.00028 \pm 0.00038
\ \ \ .
\label{eq:TMpipi}
\end{eqnarray}

\begin{figure}[!ht]
\begin{center}
\begin{minipage}[t]{8 cm}
\centerline{ \epsfig{file=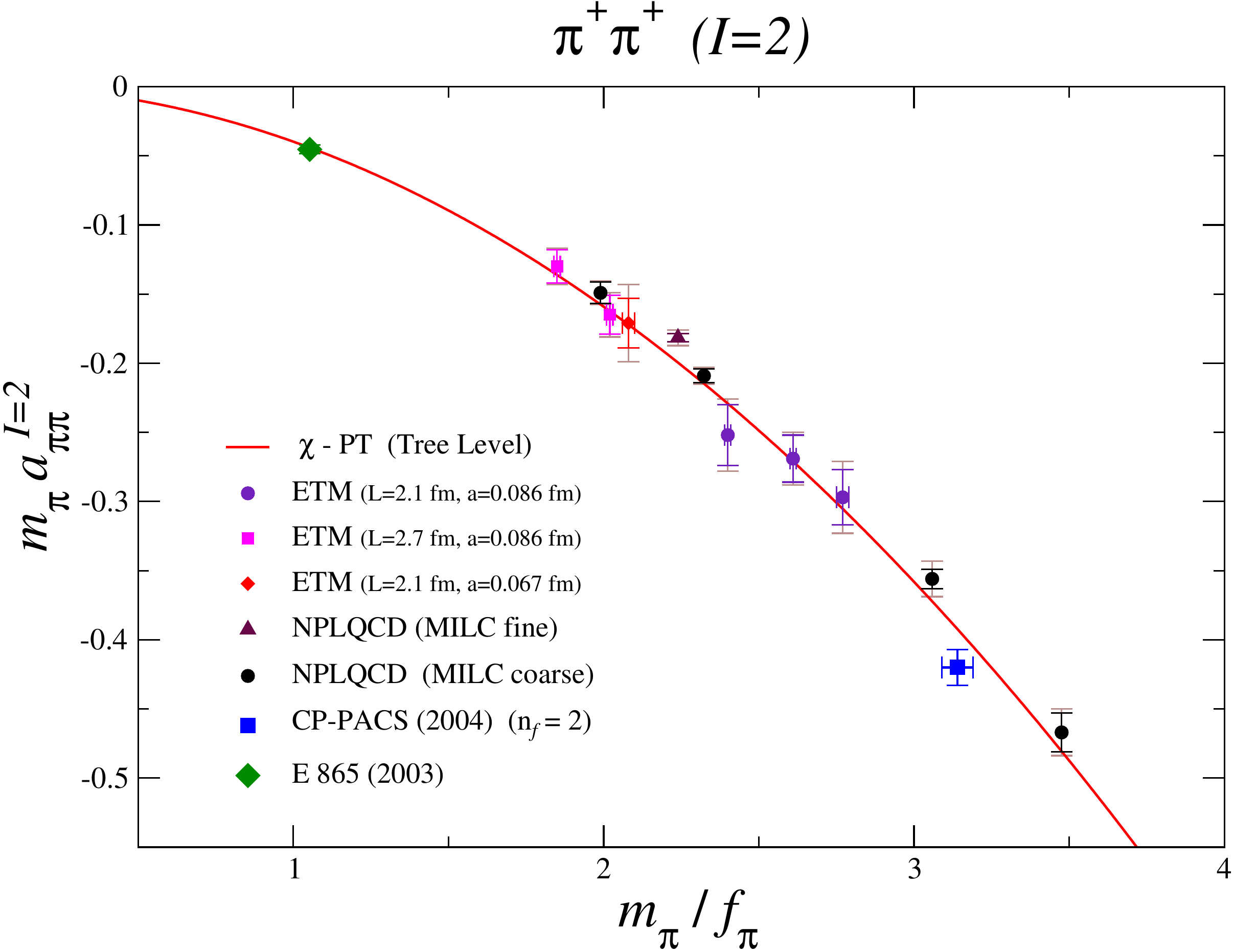,angle=0,scale=0.32}\hspace{0.54cm}\epsfig{file=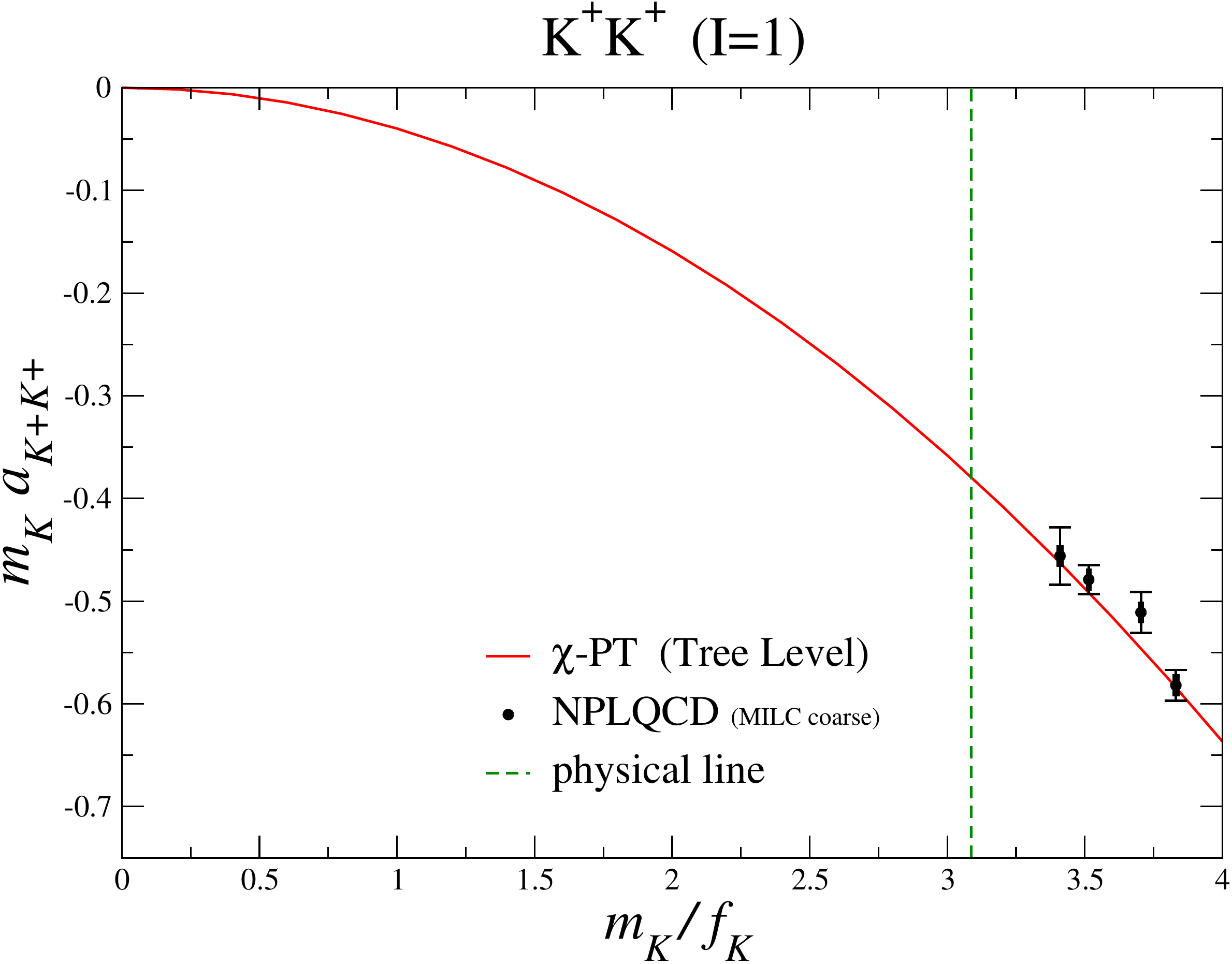,scale=0.32,angle=360}
}
\end{minipage}
\begin{minipage}[t]{16.5 cm}
\caption{$m_{\pi^+} a_{\pi^+\pi^+}$ vs. $m_{\pi^+}/f_{\pi^+}$ (left panel)
and $m_{K^+} a_{K^+K^+}$ vs. $m_{K^+}/f_{K^+}$ (right panel). 
The solid (red) curves are the current algebra predictions.
\label{fig:CAplots}}
\end{minipage}
\end{center}
\end{figure}

It is interesting to compare the pion mass dependence of the
meson-meson scattering lengths to the current algebra
predictions.  In Figure~\ref{fig:CAplots} (left panel) one sees that the
$I=2$ $\pi\pi$ scattering length is consistent with 
the current algebra result up
to pion masses that are expected to be at the edge of the chiral
regime in the two-flavor sector. While in the two flavor theory one
expects fairly good convergence of the chiral expansion and, moreover,
one expects that the effective expansion parameter is small in the
channel with maximal isospin, the lattice calculations clearly imply a
degree of 
cancellation between chiral logs and counterterms. 
However, as one sees in Figure~\ref{fig:CAplots} (right
panel), the same phenomenon occurs in $K^+K^+$ where the chiral
expansion is governed by the strange quark mass and is therefore
expected to be much more slowly converging.   This remarkable conspiracy between
chiral logs and counterterms for the meson-meson scattering lengths remains
mysterious.

\begin{figure}[!ht]
\begin{center}
\begin{minipage}[t]{8 cm}
\centerline{ \epsfig{file=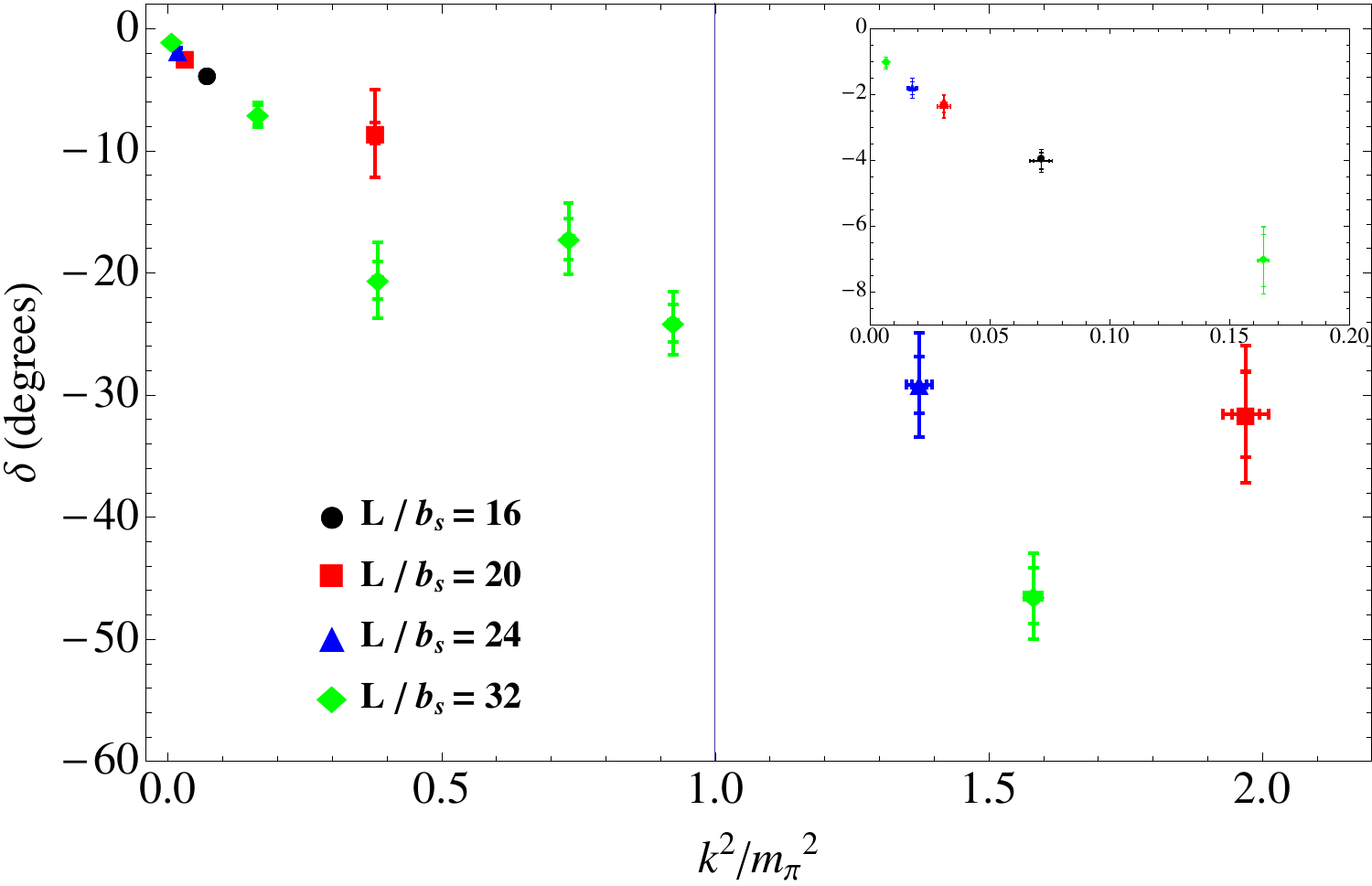 ,angle=0,scale=0.51}\hspace{0.54cm}\epsfig{file=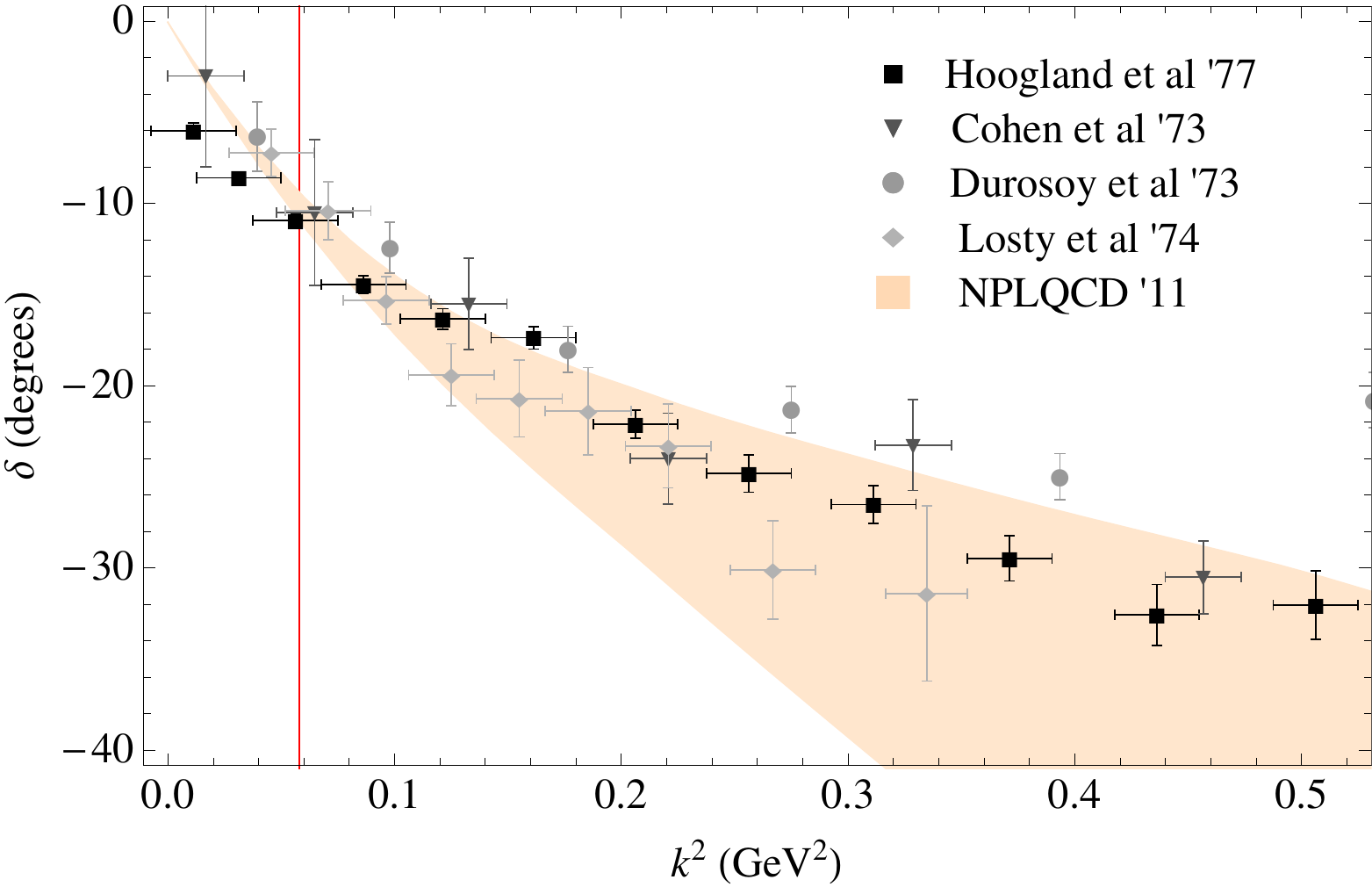,scale=0.5,angle=360}
}
\end{minipage}
\begin{minipage}[t]{16.5 cm}
\caption{
The $\pi^+\pi^+$ scattering phase-shift.  
The left panel shows the results of
the LQCD calculations below the inelastic threshold 
($|{\bf k}|^2 = 3 m_\pi^2$)
at a pion mass of
$m_\pi\sim 390~{\rm MeV}$~\protect\cite{Beane:2011sc},
obtained on anisotropic Clover
gauge-field configurations with a spatial lattice spacing of $b\sim 0.123~{\rm
  fm}$, an anisotropy of $\xi\sim 3.5$ and spatial extents of $L=16 b,20 b,24
b$ and $32 b$.  
The vertical (blue) line denotes the
start of the t-channel cut.
The shaded region in the 
right panel shows the results of the LQCD calculation extrapolated to the
physical pion mass using NLO $\chi$PT, while the points and uncertainties
corresponds to the existing experimental data.
The vertical (red) line corresponds to the inelastic threshold.
\label{fig:pipidelta}}
\end{minipage}
\end{center}
\end{figure}

LQCD calculations of the meson-meson scattering phase-shifts are much less
advanced than of the scattering length.
This is because the calculation of the phase shift, $\delta(E)$, at a
given energy, $E$, requires a lattice calculation of the two-meson correlation
function at the energy $E$.
Generally speaking, a given calculation can determine the lowest few two-hadron
energy eigenvalues for a given momentum of the center-of-mass, and  that  multiple lattice volumes
will  allow for additional  values of $E$ at which to determine $\delta(E)$.
The first serious calculation of the s-wave ($l=0$) $I=2$ $\pi\pi$ phase-shift was done by the
CP-PACS collaboration with $n_f=2$ at a relatively large pion mass~\cite{Yamazaki:2004qb}, and
recently two groups have performed calculations at lower pion 
masses~\cite{Dudek:2010ew,Beane:2011sc}, 
the results of which are shown in Figure~\ref{fig:pipidelta}.
Further, in some nice work by the Hadron Spectrum Collaboration (HSC), 
the first efforts have been made to extract 
the d-wave ($l=2$)  $I=2$ $\pi\pi$ phase shift~\cite{Dudek:2010ew}.

\subsection{\it Two-Body Bound States}

In nature, two nucleons in the $\siii-\diii$ coupled channels bind to form the 
simplest nucleus (beyond the proton), the deuteron ($J^\pi=1^+$), 
with a binding energy of $B_d = 2.224644(34)~{\rm MeV}$,
and nearly bind into a di-neutron in the $\si$ channel.
However, little is known experimentally about possible bound states in more
exotic channels, for instance those containing strange quarks.  
The most famous exotic channel that has been postulated to support a bound
state (the  H-dibaryon~\cite{Jaffe:1976yi})
has the quantum numbers of $\Lambda\Lambda$ (total angular momentum
$J^\pi=0^+$, isospin $I=0$ and strangeness $s=-2$).  
In this channel, all six quarks in naive quark models, like the MIT bag model,
can be in the lowest-energy single-particle state.
Additionally, more extensive analyses using one-boson-exchange (OBE) models~\cite{Stoks:1999bz} and
low-energy effective field theories
(EFT)~\cite{Miller:2006jf,Haidenbauer:2009qn}, 
both constrained by experimentally
measured  nucleon-nucleon (NN) and  hyperon-nucleon (YN) cross-sections 
and the approximate SU(3) flavor symmetry of
the strong interactions, suggest that other exotic channels also support
bound states.  
In the limit of SU(3) flavor symmetry, the $\si$-channels
are in symmetric irreducible representations of
${\bf 8}\otimes {\bf 8} = {\bf 27}\oplus {\bf 10}\oplus \overline{{\bf 10}}
\oplus {\bf 8} \oplus {\bf 8} \oplus {\bf 1}$, and hence the $\Xi^-\Xi^-$,
$\Sigma^-\Sigma^-$, and $nn$ 
(along with $n\Sigma^-$ and $\Sigma^-\Xi^-$)
are in the ${\bf 27}$.
YN and NN scattering data, along with the leading SU(3) breaking effects
from the light-meson and baryon masses, suggest that $\Xi^-\Xi^-$  and
$\Sigma^-\Sigma^-$  are bound at the physical values of the light-quark 
masses~\cite{Stoks:1999bz,Miller:2006jf,Haidenbauer:2009qn}.

\begin{figure}[!ht]
\begin{center}
\begin{minipage}[t]{8 cm}
\centerline{\epsfig{file=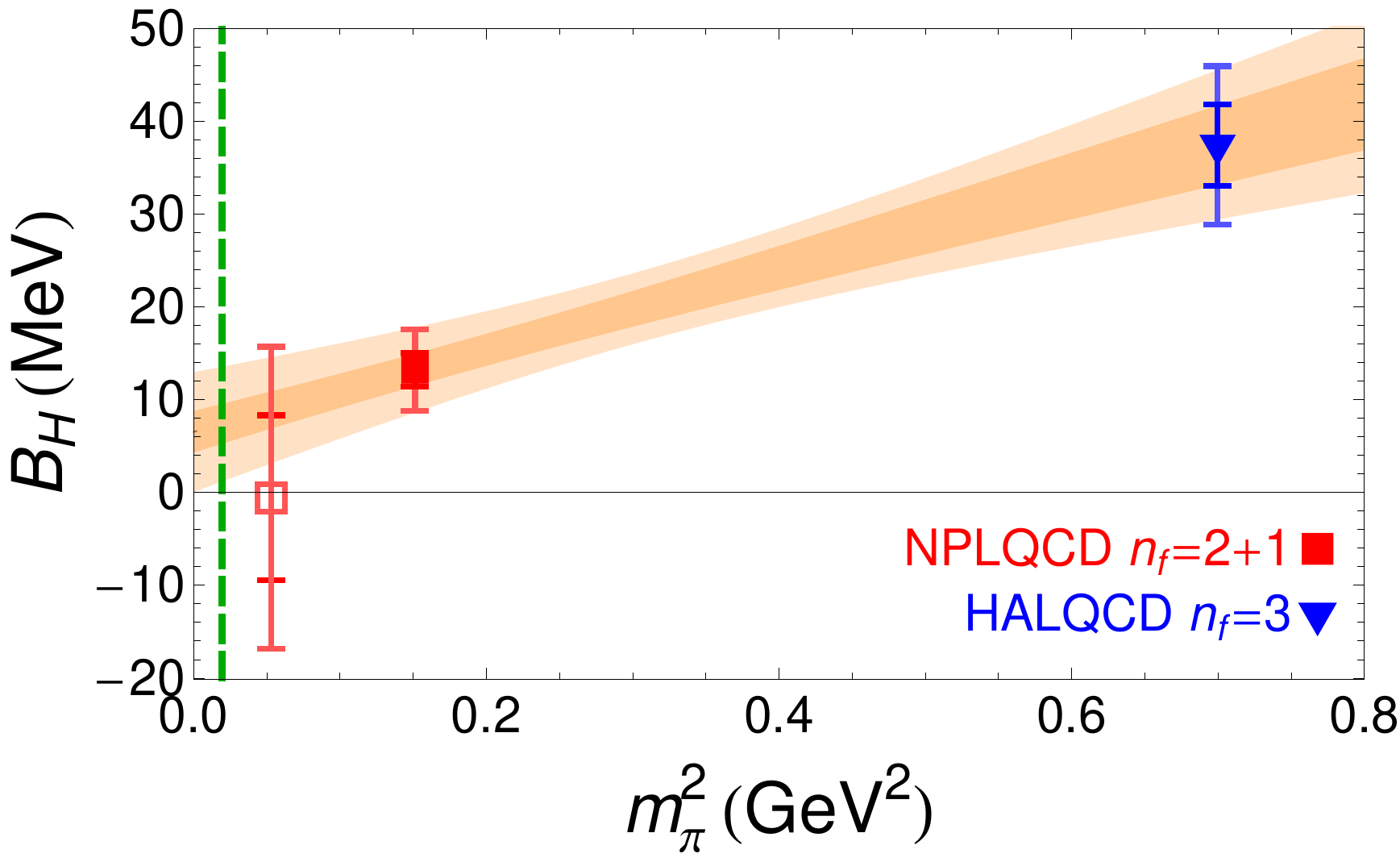,scale=0.5}\epsfig{file=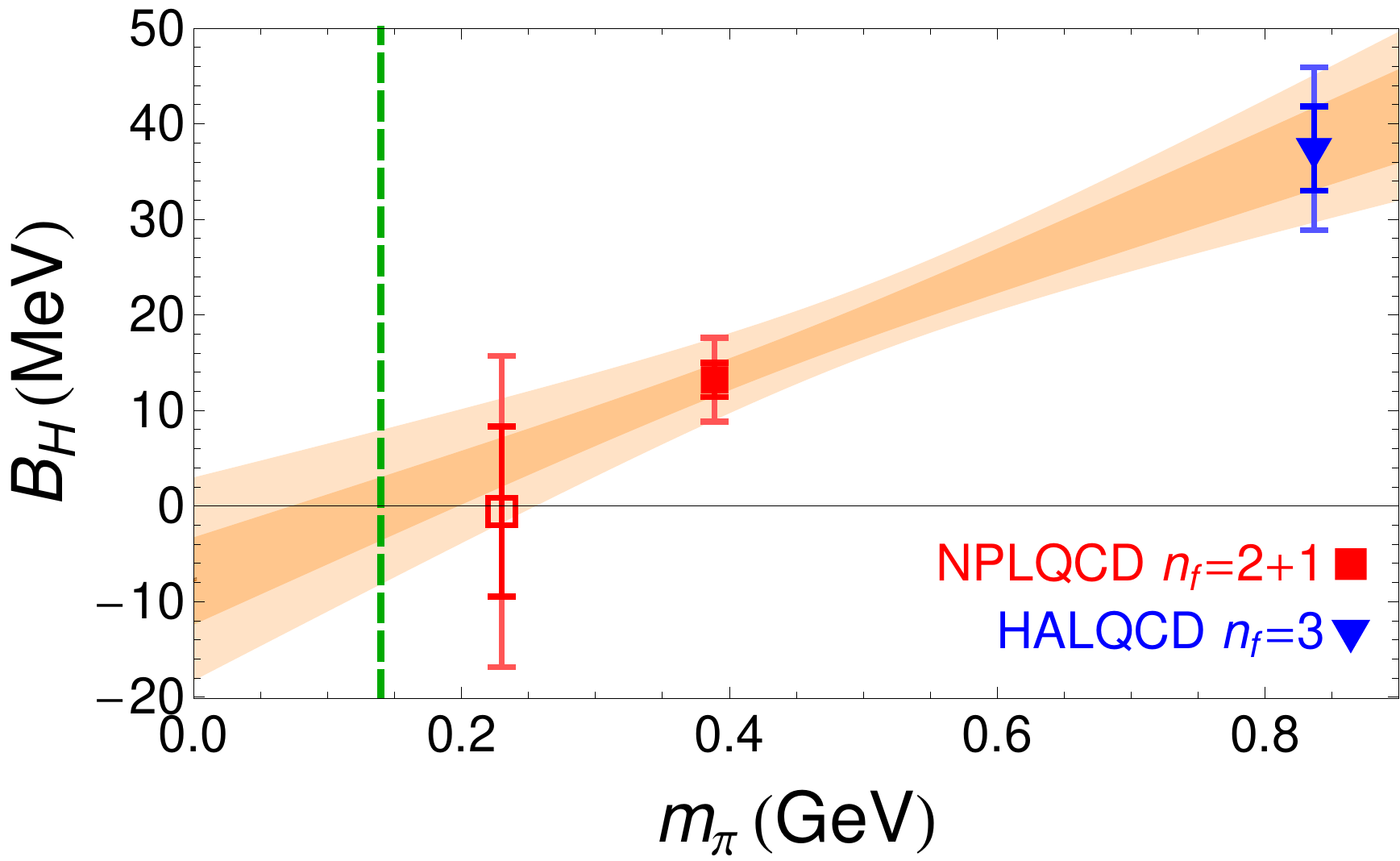,scale=0.5}}
\end{minipage}
\begin{minipage}[t]{16.5 cm}
\caption{Possible extrapolations of the LQCD results for the binding of the H-dibaryon.
The left panel corresponds to an extrapolation that is quadratic in $m_\pi$, of
the form $B_H(m_\pi) = B_0 + d_1 m_\pi^2$, while the right panel corresponds to
an extrapolation of the form $B_H(m_\pi) = \tilde B_0 +
\tilde d_1 m_\pi$.
In each panel, 
the red points are the result from the $n_f=2+1$ calculations of 
NPLQCD~\protect\cite{Beane:2010hg,Beane:2011iw}
while the blue point is from the $n_f=3$
calculation of HALQCD~\protect\cite{Inoue:2010es}.
The green dashed vertical line corresponds to the physical pion mass.
 \label{fig:Hextrap}}
\end{minipage}
\end{center}
\end{figure}

Recently, $n_f=2+1$ calculations~\cite{Beane:2010hg,Beane:2011iw},  
and subsequent $n_f=3$
calculations~\cite{Inoue:2010es}, 
have provided evidence that the H-dibaryon  
is bound at a pion mass of $ m_\pi\sim 390~{\rm MeV}$ [NPLQCD]
and
$m_\pi\sim 837~{\rm MeV}$ [HALQCD]~\footnote{Both
  calculations were performed at approximately the same spatial lattice spacing
of $b\sim 0.12~{\rm fm}$.}. 
The infinite-volume extrapolated H-dibaryon binding energy at $ m_\pi\sim 390~{\rm MeV}$
is found to be
\begin{eqnarray}
B_{H}^{(L=\infty)} & = & 
13.2\pm 1.8 \pm 4.0~{\rm MeV}
\ \ \ .
\label{eq:HbindingLQCDextrap}
\end{eqnarray}
Possible extrapolations to the physical
light-quark masses 
are shown in Figure~\ref{fig:Hextrap}, and 
suggest a weakly bound H-dibaryon or a near threshold
resonance exists in this channel~\cite{Beane:2011xf,Shanahan:2011su,Haidenbauer:2011ah}.

\begin{figure}[!ht]
\begin{center}
\begin{minipage}[t]{8 cm}
\centerline{\epsfig{file=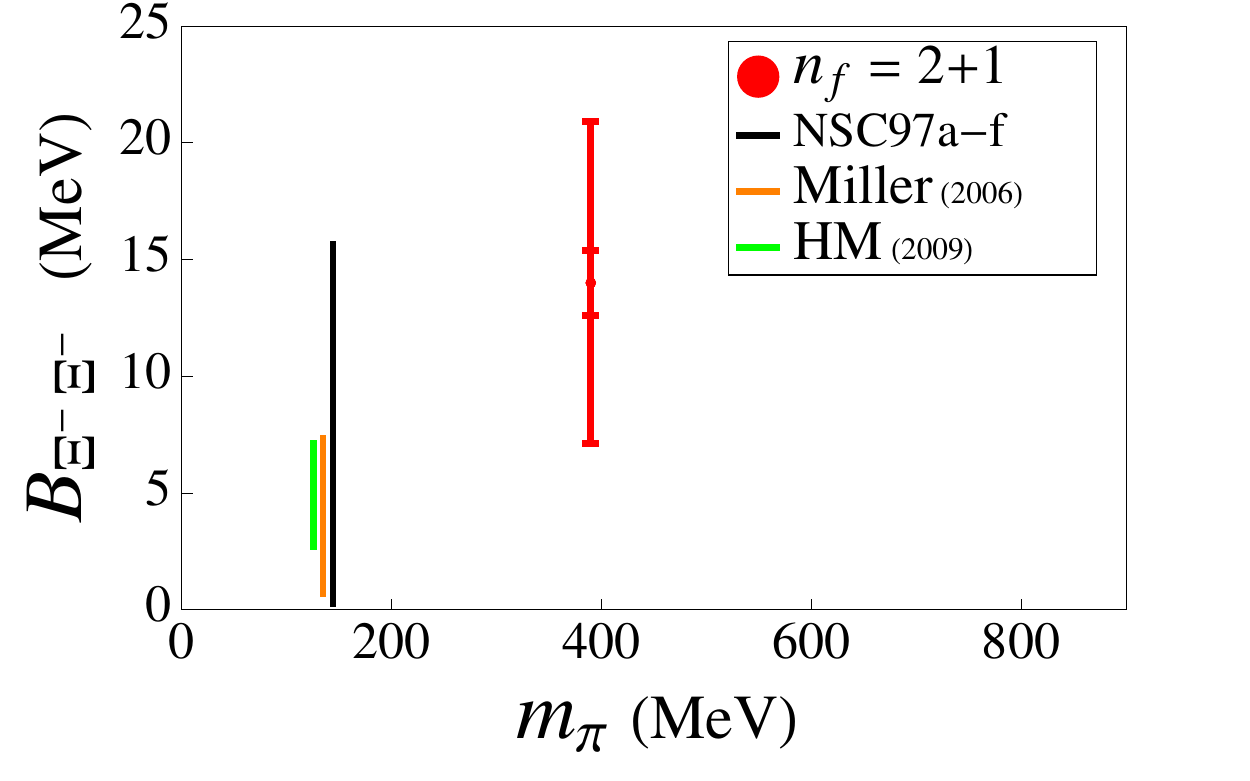,scale=0.85}}
\end{minipage}
\begin{minipage}[t]{16.5 cm}
\caption{The $\Xi^-\Xi^-$ binding energy as a function of the pion mass.
The black line denotes the predictions of the NSC97a-NSC97f  
models~\protect\cite{Stoks:1999bz} constrained by
nucleon-nucleon and hyperon-nucleon scattering data.
The orange line denotes the range of predictions by Miller~\cite{Miller:2006jf},
and the green line denotes the leading order EFT prediction by Haidenbauer and
Meissner (HM)~\cite{Haidenbauer:2009qn}.
The red point and uncertainty 
(the inner is statistical and the outer is statistical and systematic
combined in quadrature)
is the NPLQCD $n_f=2+1$ result~\protect\cite{Beane:2011iw}.
The OBE model and EFT predictions at the physical pion mass are displaced horizontally for
the purpose of display.
 \label{fig:XiXiALL}}
\end{minipage}
\end{center}
\end{figure}

The NPLQCD collaboration has also found evidence that the $\Xi^-\Xi^-$-system
is bound at a pion mass of $ m_\pi\sim 390~{\rm MeV}$~\cite{Beane:2011iw}, 
as shown in 
Figure~\ref{fig:XiXiALL}. 
This result, and the predictions of OBE models and leading order (LO) EFT, 
are shown in
Figure~\ref{fig:XiXiALL}.
It is important to note that the uncertainty of the LQCD
result is comparable to that of the OBE models and EFT results,
and
demonstrates that LQCD is approaching the time
where it will provide more precise constraints on exotic systems
than currently possible in the laboratory.
It will be interesting to see whether J-PARC~\cite{jparc} 
or FAIR~\cite{Steinheimer:2008hr}
can
provide constraints on the $s=-3$ and $s=-4$ systems, 
as well as on the H-dibaryon~\cite{jparcHdib}.
The calculated binding energy of
\begin{eqnarray}
B_{\Xi^-\Xi^-}^{(L=\infty)} & = & 
14.0\pm 1.4 \pm 6.7~{\rm MeV}
\ \ \ ,
\label{eq:XiXibindingLQCDextrap}
\end{eqnarray}
at $ m_\pi\sim 390~{\rm MeV}$ provides strong motivation to return to OBE models and EFT frameworks
and determine the expected dependence on the light-quark masses.
The same LQCD calculations also found
hints that both the deuteron and the di-neutron are bound at this same pion
mass, with binding energies of 
\begin{eqnarray}
B_d^{(L=\infty)} & = & 
11\pm 5\pm 12~{\rm MeV}
\ \ \ , \ \ \ 
B_{nn}^{(L=\infty)} \ = \ 
7.1\pm 5.2\pm 7.3~{\rm MeV}
\ \ .
\label{eq:deutbindingLQCDextrap}
\end{eqnarray}
\begin{figure}[!ht]
\begin{center}
\begin{minipage}[t]{8 cm}
\centerline{\epsfig{file=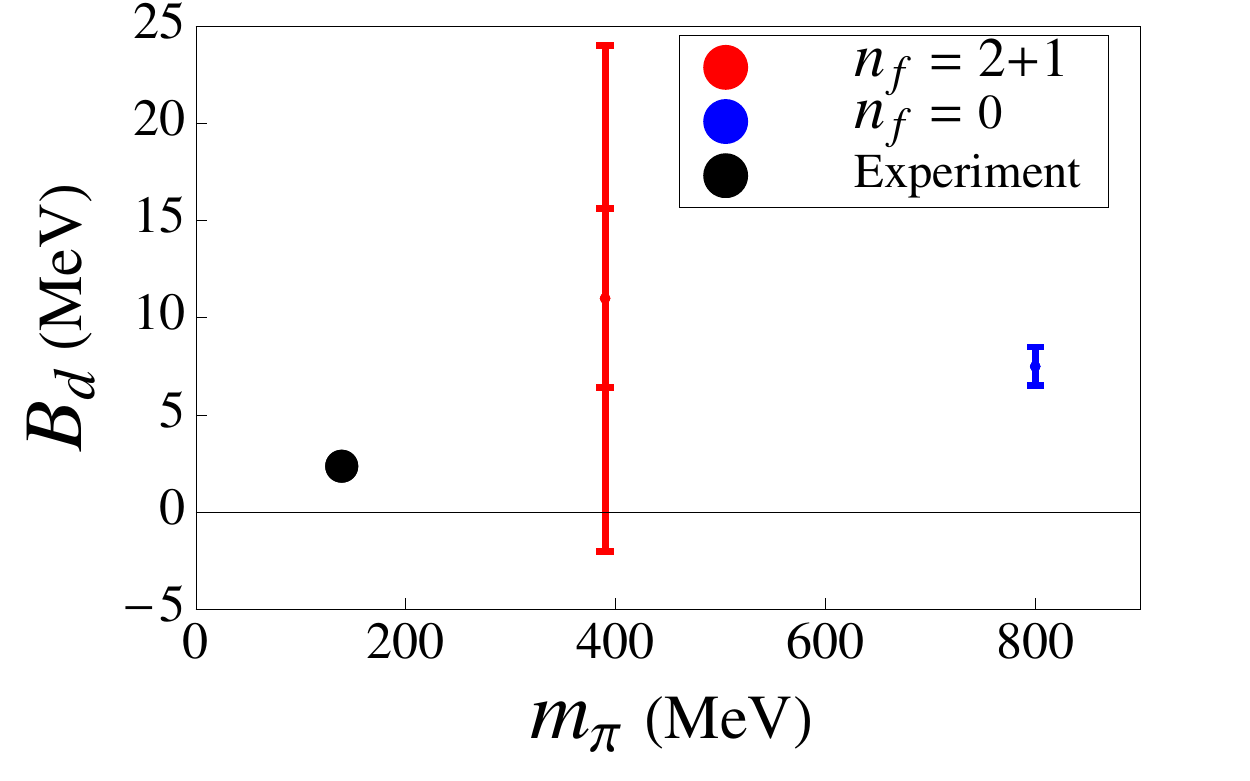,scale=0.7}\epsfig{file=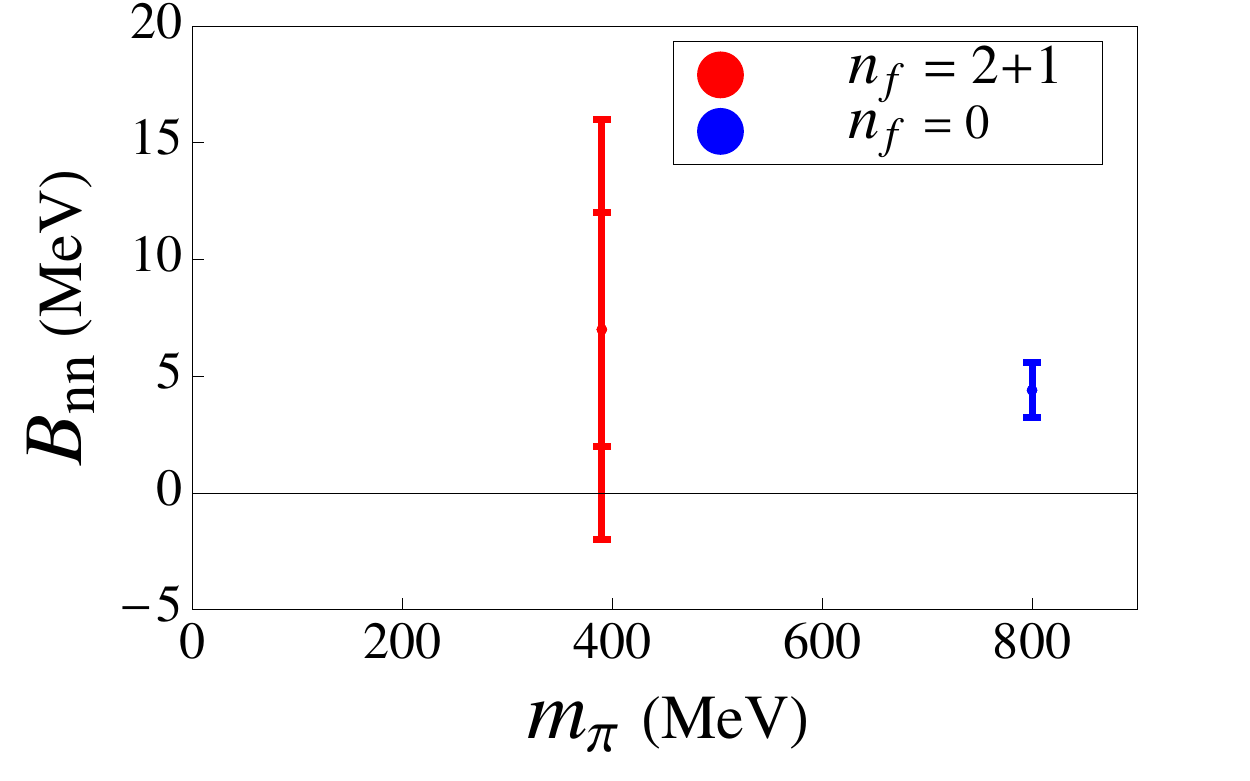,scale=0.7}}
\end{minipage}
\begin{minipage}[t]{16.5 cm}
\caption{The deuteron (left panel) and di-neutron (right panel) binding
  energies 
as a function of the pion mass.
The black circle denotes the experimental value.
The blue points and uncertainties are  from the quenched
calculations of Ref.~\protect\cite{Yamazaki:2011nd}, while the red points and
uncertainties 
(the inner is statistical and the outer is statistical and systematic
combined in quadrature)
are from the NPLQCD $n_f=2+1$ calculations~\protect\cite{Beane:2011iw}.
 \label{fig:dnnALL}}
\end{minipage}
\end{center}
\end{figure}
The results of these calculations, along with the results of 
recent quenched calculations~\cite{Yamazaki:2011nd} are shown in Figure~\ref{fig:dnnALL}.

\begin{figure}[tb]
\begin{center}
\begin{minipage}[t]{8 cm}
\centerline{\epsfig{file=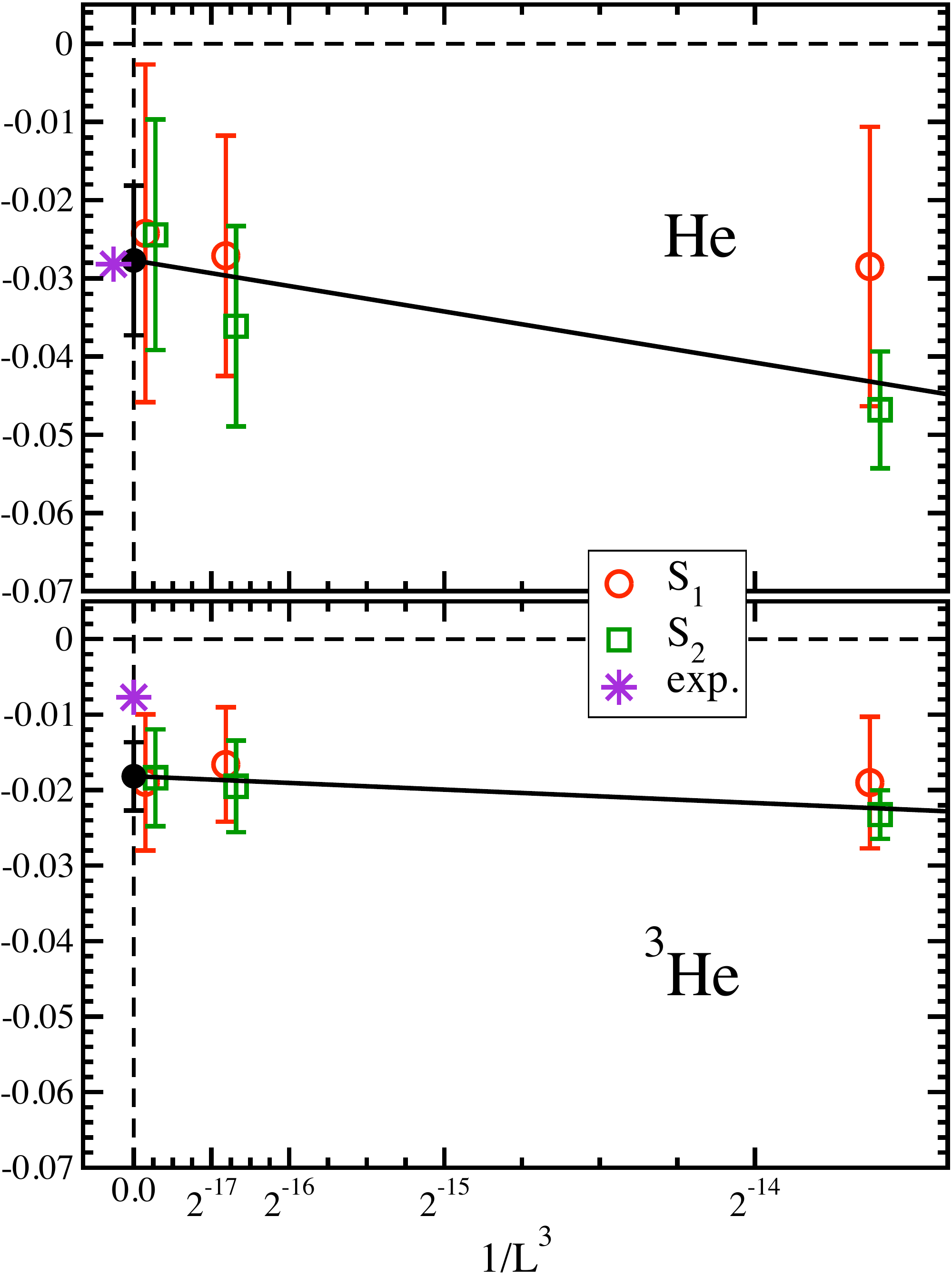,scale=0.35}}
\end{minipage}
\begin{minipage}[t]{16.5 cm}
\caption{
Quenched results 
for the binding energies (in lattice units)
in the triton
channel (lower panel) and 
the $\alpha$-particle
channel (upper panel) obtained by the PACS-CS collaboration~\cite{Yamazaki:2009ua}. 
The pion mass is
$m_\pi\sim 800~{\rm MeV}$.
[Image is reproduced with the permission of the PACS-CS collaboration.]
\label{fig:alpha}}
\end{minipage}
\end{center}

\end{figure}

In late 2009,
the PACS-CS collaboration performed the first quenched calculation of 
a four-baryon correlation function~\cite{Yamazaki:2009ua} 
in the $\alpha$-particle ($^4$He nucleus) channel.
The pion mass was $m_\pi\sim 800~{\rm MeV}$
and sea quark effects were ignored, however, 
this is a very important step towards calculating the 
properties and interactions of nuclei.  
Their results are shown in
Figure~\ref{fig:alpha},
along with their result in the triton channel,
and 
significantly improved
statistics are hoped for in the near future.
%

\subsection{\it Nucleon-Nucleon Interactions}

Perhaps the most studied and best understood of the two-hadron processes
are proton-proton and proton-neutron scattering.  In the s-wave, only two
combinations of spin and isospin are possible, the spin-triplet
isosinglet $np\ (^3S_1)$ and the spin-singlet isotriplet $pp\ (^1S_0)$, 
$np\ (^1S_0)$ and $nn\ (^1S_0)$.
At the physical pion mass, the scattering lengths in these channels
are unnaturally large and the $\siii-\diii$ coupled-channel contains 
the deuteron.
These large scattering lengths and the deuteron are
described in EFT, by the coefficient of the
momentum-independent four-nucleon operator being near  a non-trivial
fixed-point~\cite{Kaplan:1998tg,Birse:1998dk} in its renormalization group flow
at the physical light-quark masses.  An
interesting line of investigation is the study of the scattering
lengths as a function of the quark masses to ascertain the sensitivity
of this fine-tuning to the QCD
parameters~\cite{Beane:2002vs,Beane:2002xf,Epelbaum:2002gb}.
While  the fine
tuning is not expected to persist away from the physical masses, it is
interesting to determine how the structure of nuclei depend upon the
fundamental constants of nature.

The first study of baryon-baryon scattering with LQCD was performed more
than a decade ago by Fukugita~{\it et
al}~\cite{Fukugita:1994na,Fukugita:1994ve}.  Those calculations were
quenched and at relatively large pion masses, $m_\pi\gsim 550~{\rm
MeV}$.  Since that time, the dependence of the NN scattering lengths upon the
light-quark masses has been determined to various non-trivial orders
in the EFT expansion~\cite{Beane:2002vs,Beane:2002xf,Epelbaum:2002gb},
which are estimated to be valid up to $m_\pi\sim 350~{\rm
MeV}$. Therefore, predictions of NN scattering parameters becomes
possible with LQCD calculations that are performed
with $m_\pi \lsim 350~{\rm MeV}$.

\begin{figure}[!ht]
\begin{center}
\begin{minipage}[t]{8 cm}
\centerline{\epsfig{file=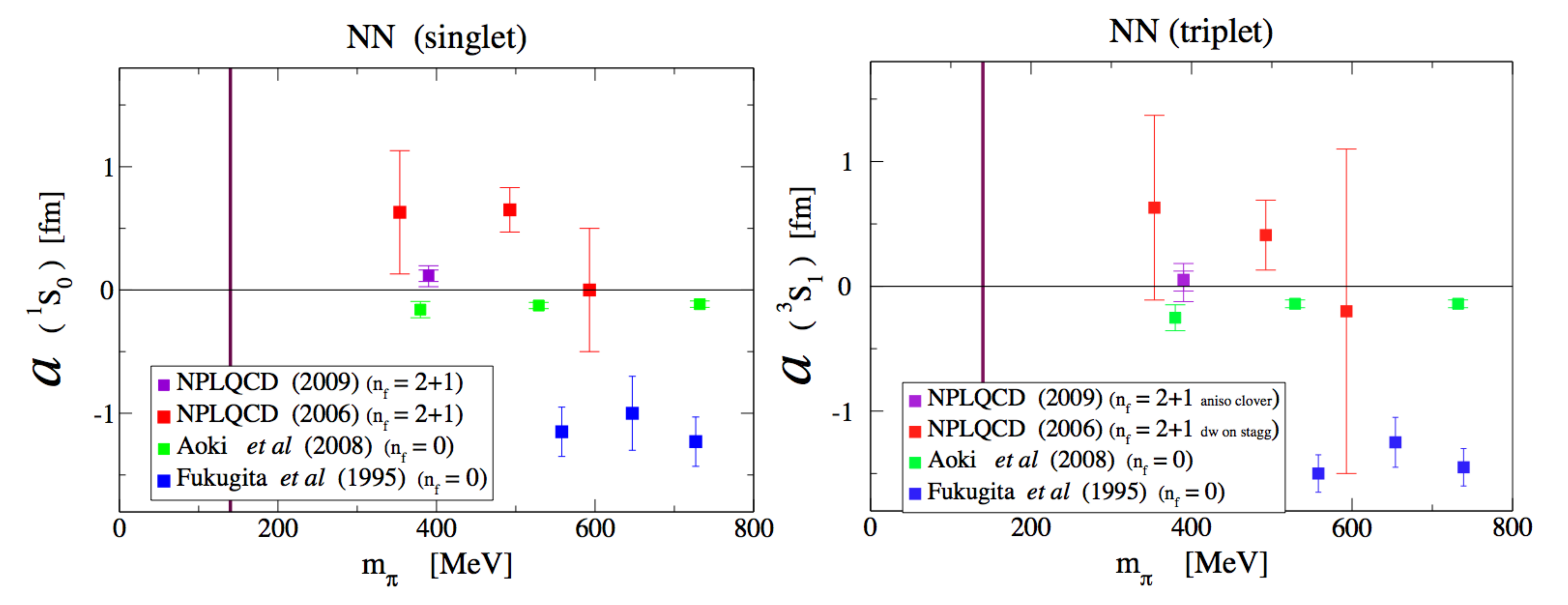,scale=0.3}}
\end{minipage}
\begin{minipage}[t]{16.5 cm}
\caption{A compilation of the scattering lengths for NN scattering
    in the $^1S_0$ channel (left panel) and $^3S_1$ channel 
(right panel) calculated
    with LQCD~\cite{Beane:2006mx,Beane:2009py} 
and with quenched LQCD~\protect\cite{Aoki:2008hh,Aoki:2009ji,Ishii:2006ec}.  
The vertical lines correspond to the physical pion mass.
\label{fig:NN-ALL-LQCD}}
\end{minipage}
\end{center}
\end{figure}

The NPLQCD collaboration performed the first $n_f=2+1$ LQCD calculations
of nucleon-nucleon interactions~\cite{Beane:2006mx} and
hyperon-nucleon~\cite{Beane:2006gf} interactions at low-energies but with
unphysical pion masses, and  the nucleon-nucleon scattering lengths were found
to be of natural size. 
The fine-tunings at the physical values of the light-quark masses indicate
that  LQCD calculations with quark masses much
closer to the physical values (than today) 
are needed to extrapolate to the experimental
values. The results of the LQCD calculation at the lightest
pion mass and the experimentally-determined scattering lengths at the
physical value of the pion mass were used to constrain the chiral
dependence of the scattering lengths from $m_\pi\sim 350~{\rm MeV}$
down to the chiral limit~\cite{Beane:2006mx}. 
However, these results suggest various
possible scenarios toward the chiral limit
which can only be resolved by way of LQCD
calculations at lighter pion masses.  
Summaries of existing LQCD calculations of the NN scattering lengths are shown
in Figure~\ref{fig:NN-ALL-LQCD}.
In contrast, very little is
known about the interactions between nucleons and hyperons from
experiment, and future LQCD calculations will likely provide the best
determinations of the corresponding scattering parameters and hence
determine the role of hyperons in neutron stars.
%

\subsection{\it Meson Condensates}

\begin{figure}[!ht]
   \centering
\begin{minipage}[!ht]{8 cm}
\centerline{
  \epsfig{file=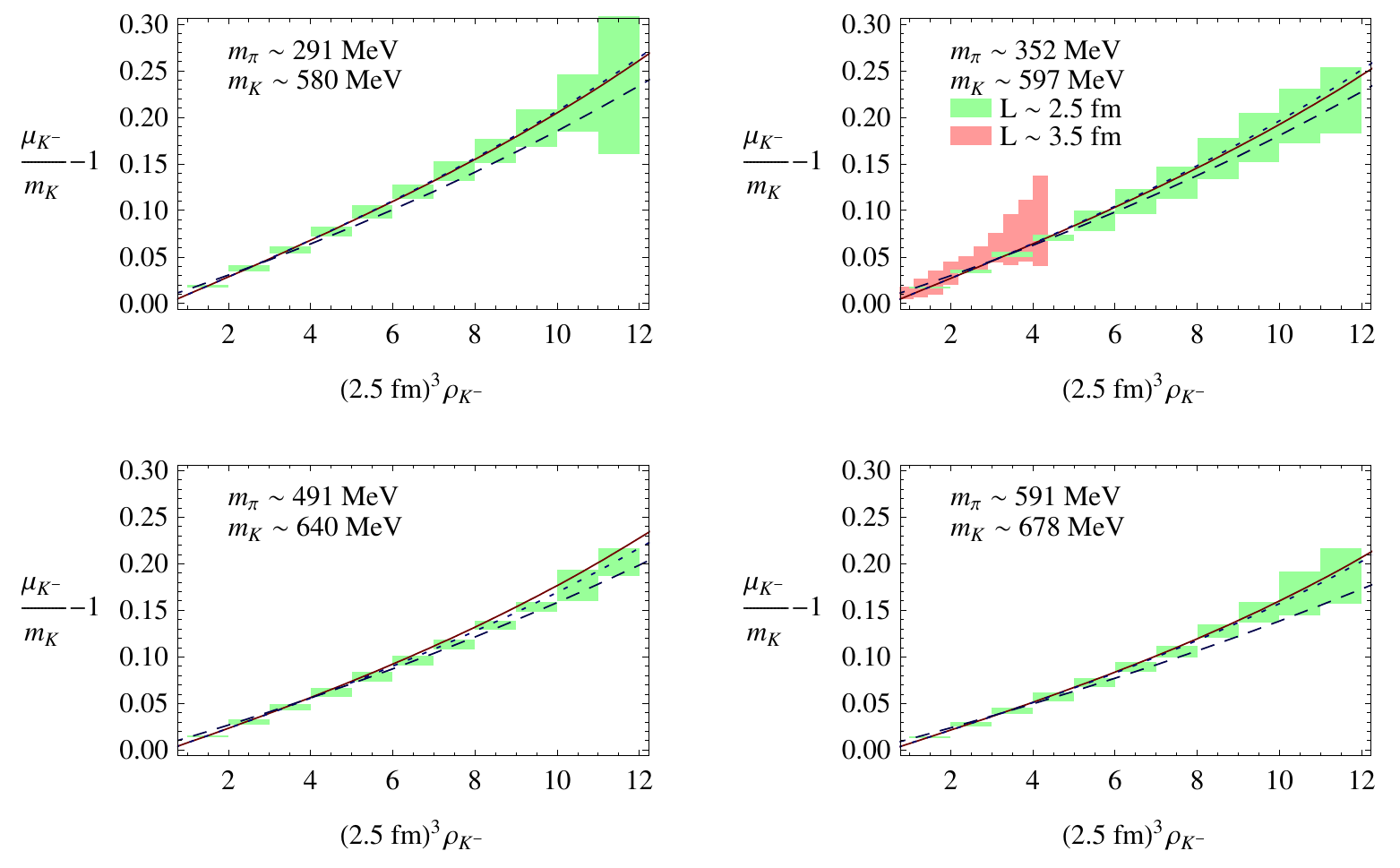,scale=0.99} }
\end{minipage}
\begin{minipage}[t]{16.5 cm}
   \caption{The dependence of the
     strangeness chemical potential on the kaon
     density~\protect\cite{Detmold:2008yn}. The curves
     correspond to the predictions of tree level chiral perturbation
     theory (dashed)~\protect\cite{Son:2000xc}, the fitted energy shift
     (solid) and without the three-body interaction (dotted).}
   \label{fig:str}
\end{minipage}
 \end{figure}

The ground state of a generic system of many bosons with weak repulsive
interactions is a Bose-condensate. 
A QCD system of pions and/or
kaons  
form a Bose-Einstein condensate with  fixed third-component 
of isospin, $I_z$, and strangeness, $s$, and  it is of
significant theoretical and phenomenological interest to investigate
the properties of such systems. Theoretical efforts have used 
LO $\chi$-PT
to investigate the phase diagram at
low chemical potential~\cite{Son:2000xc} and it is important to assess
the extent to which these results agree with QCD, because in neutron stars, it
is possible that it is energetically favorable to (partially-)
electrically neutralize the
system with a condensate of $K^-$ mesons instead of electrons.
Numerical calculations provide  a probe of
the dependence of
the energy on the pion  or kaon density, and thereby allow for 
an extraction of the
chemical potential via a finite difference~\cite{Detmold:2008yn}. 
Results  from mixed-action calculations of kaon condensates
are shown in Figure~\ref{fig:str}~\cite{Detmold:2008yn}, 
along with the predictions of tree-level $\chi$-PT, 
which are in remarkably good agreement.
This is encouraging for studies of kaon condensation in
neutron stars where, typically, tree level $\chi$-PT
interactions  amongst kaons, and between kaons and 
baryons~\cite{KaplanNelson}, are assumed.
These calculations have been extended to mixed kaon-pion systems in Ref.~\cite{Detmold:2011kw}.

\section{The Next Decade - the Drive Toward the Exa-scale}

\begin{figure}[tb]
\begin{center}
\begin{minipage}[t]{8 cm}
\centerline{\epsfig{file=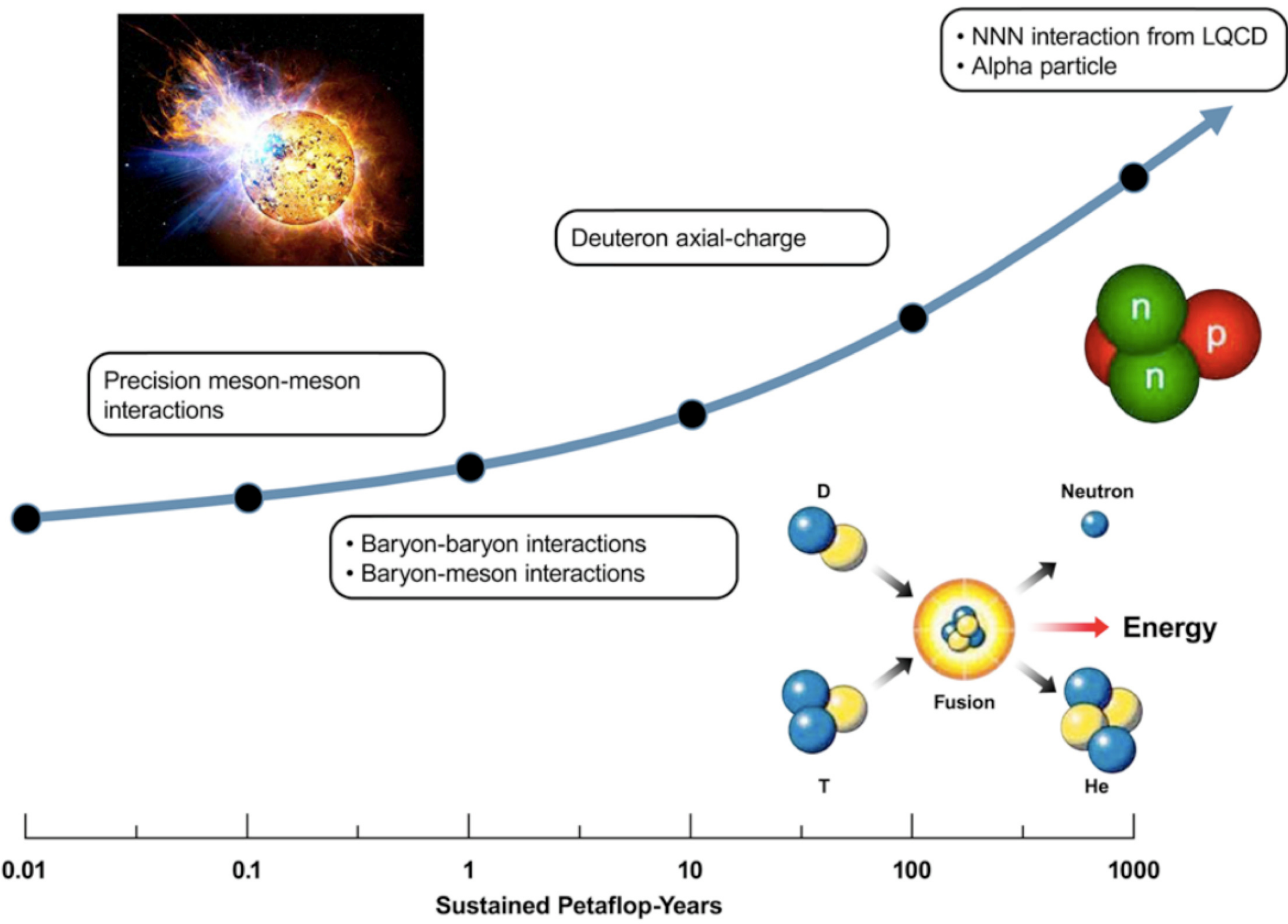,scale=0.55}}
\end{minipage}
\begin{minipage}[t]{16.5 cm}
\caption{
Estimates of the resources required to complete calculations of importance
to nuclear physics~\protect\cite{ExascaleReport:2009}. 
Except for the quantities indicated as requiring exa-scale
resources, 
the resource requirements are for 
calculations performed in the isospin limit without
electroweak interactions. 
\label{fig:BBresources}}
\end{minipage}
\end{center}
\end{figure}

A number of workshops focusing on the science need for exa-scale
computing resources sponsored  by the US Department of Energy were held during
2009.
One of the workshops, {\it Forefront Questions in Nuclear Science and the Role of
  Computing at the Extreme Scale}~\cite{ExascaleReport:2009}, 
established the need for exa-scale computing
resources in order for the main goals of the field of nuclear physics to be 
accomplished~\cite{Savage:2010hp}.
One of the major goals of the field that requires exa-scale computing
resources is the calculation of nuclear forces from QCD using Lattice QCD,
and Figure~\ref{fig:BBresources} presents 
an overview of
current estimates of these requirements.

As reviewed in Refs.~\cite{Beane:2008dv,Beane:2010em}, 
a complete calculation of the nucleon-nucleon
scattering amplitude, and the hyperon-nucleon and hyperon-hyperon
scattering amplitudes 
(including multiple lattice spacings, volumes and light quark masses)
will require sustained peta-scale resources, as shown in 
Figure~\ref{fig:BBresources}.
The same is true for the meson-baryon interactions.  
It is estimated that sustained sub-peta-flop-year resources are
required to perform high-precision calculations of meson-meson
scattering-phase  shifts, including the
contributions from disconnected diagrams to the isosinglet $\pi\pi$ channel.
Further, it is estimated that sustained peta-scale resources are required in
order to calculate the matrix elements of electroweak operators, such as those
determining neutrino-induced breakup of the deuteron, in the few-nucleon
sector.
Significant progress is being made in computing single hadron matrix elements of such
operators, such as the isovector axial-current matrix element in the nucleon,
$g_A$~\cite{Khan:2006de,Bratt:2010jn,Yamazaki:2009zq}.  
While the extrapolation to the physical pion mass, and to infinite
volume remain the subject of discussions in the community, relatively rapid
progress is being made.  Calculations of matrix elements of operators that
receive contributions from disconnected diagrams remain difficult with
currently available resources, but will be addressed with peta-scale resources.
A significant  uncertainty
in the experimentally determined  properties of neutrinos comes from
the uncertainties in weak matrix elements between nuclear states.
Such  uncertainties in few-nucleon systems should be reduced within
the next decade with anticipated  Lattice QCD calculations, as indicated in 
Figure~\ref{fig:BBresources}.

Despite the first Lattice QCD calculations of three-
and
four-baryon systems
appearing recently, it is estimated that exa-scale computing resources will be
required to extract precision nuclear interactions among three-nucleons and
determine the spectrum of the $\alpha$-particle.
Given that the three-nucleon interaction is relatively imprecisely known 
from experiment
when
compared with the two-nucleon interactions, this calculation will have
significant impact upon nuclear structure and reaction calculations.  
The three-baryon interactions between strange and non-strange baryons will be
calculable with the same level of precision
with minimal additional resource requirements.

Current discussions regarding exa-scale computing facilities suggest that
it may be possible to see such resources deployed sometime around 
2018~\cite{ExascaleReport:2009}.
Clearly, such resources are required for the calculation of quantities of
central importance to the nuclear physics program. 
During the next decade the field will develop the ability to perform low-energy 
strong interaction calculations with quantifiable uncertainty estimation.

\section{Formal Developments}

\begin{figure}[tb]
\begin{center}
\begin{minipage}[t]{8 cm}
\centerline{\epsfig{file=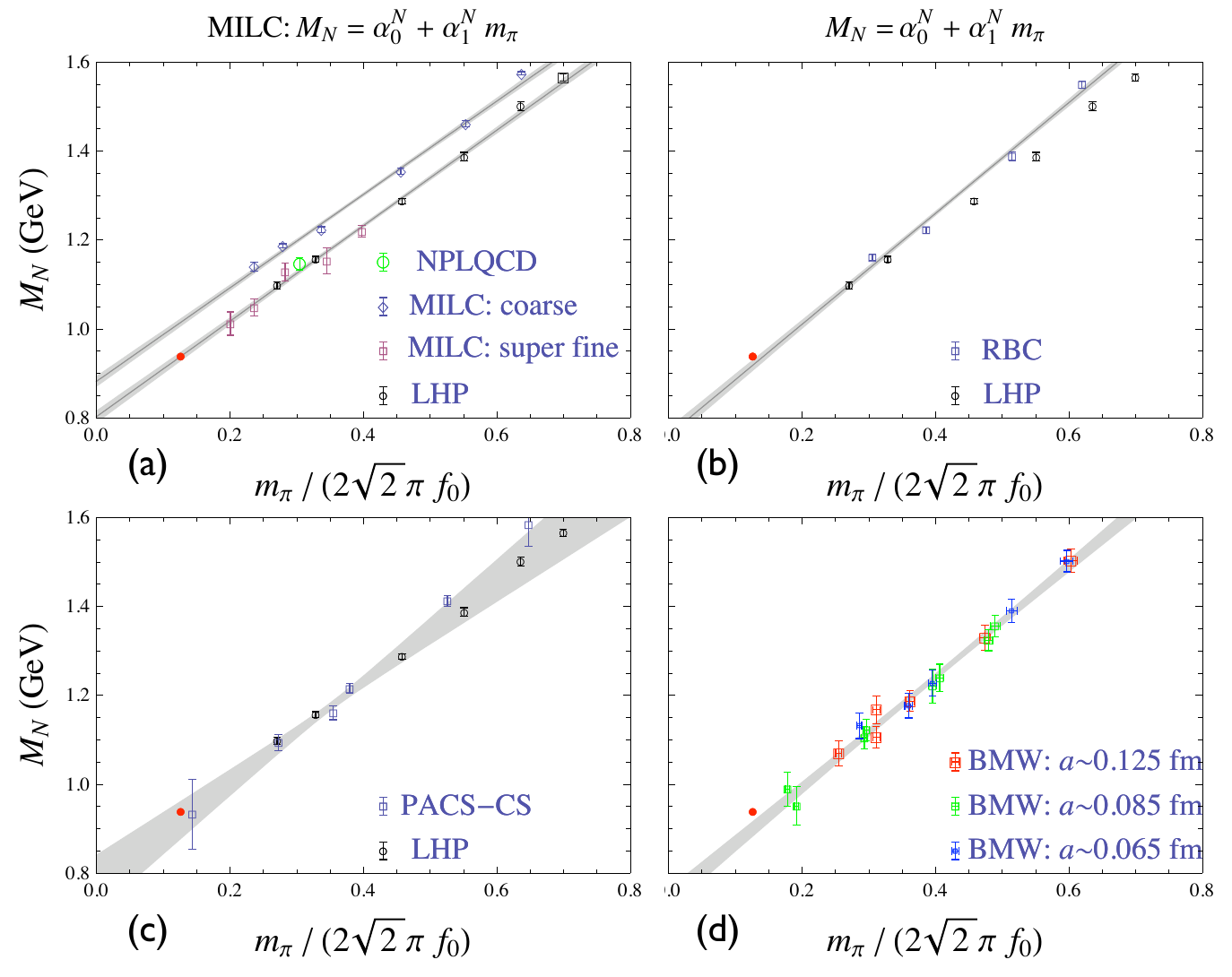,scale=0.95}}
\end{minipage}
\begin{minipage}[t]{16.5 cm}
\caption{
A summary of LQCD calculations of the mass of the 
nucleon (as of 2009)~\protect\cite{WalkerLoud:2008pj}.
[Image is reproduced with the permission of A. Walker-Loud.]
}
\label{fig:MN}
\end{minipage}
\end{center}
\end{figure}

In addition to the large computational resources and algorithmic
improvements that are required to complete the mission, 
formal developments are also required. 
For instance, a reliable method with which to extract inelastic
scattering cross-sections from LQCD calculations does not yet exist, and 
must be developed.
Further, a better understanding of the convergence of EFT's that are 
used in such systems is required.  Current LQCD calculations indicate 
unexpected
behaviors in the convergence patterns of the EFT's, including chiral 
perturbation
theory ($\chi$PT). 
Two striking examples of such behaviors are evident in LQCD calculations of the
nucleon mass, exhibiting a light-quark mass dependence that is consistent with 
being linear in the pion mass (i.e. $\sqrt{m_q}$)~\cite{WalkerLoud:2008pj}, 
as shown in Figure~\ref{fig:MN},
and in the meson-meson scattering lengths which are consistent with their 
tree-level
predictions, even at large light-quark masses~\cite{Beane:2007xs}.
%

\section{Outlook}

A central goal of the field of nuclear physics is to establish a framework with
which to perform high-precision calculations, with quantifiable uncertainties,
of strong-interaction processes occurring under a broad range of conditions.
Quantum chromodynamics was established as the underlying theory of the strong
interactions during the 1970's, however, nuclear physics is 
in the regime of QCD in which its defining property of 
asymptotic freedom is hidden by the vacuum and by 
the phenomenon of confinement.
Lattice QCD, in which the QCD path-integral is evaluated numerically, is the
only known way to perform rigorous  QCD calculations of low-energy
strong interaction processes.
With the research into, and development of, high-performance computing, nuclear
physics,  quantum-field theory, applied
mathematics, and numerical algorithms  that has taken
place over the last few decades, the field of nuclear physics is entering into
an era in which Lattice QCD will become a quantitative tool in much the same
way that experiments are, but with a different scope and different range of
applicability.
Rapid progress is currently being made in the calculation of the interactions
among the low-lying baryons. 
Present day Lattice QCD calculations are being performed at
pion masses larger than the physical pion mass, but as exploratory 
calculations are
now being performed at the physical pion mass, the interactions among  baryons
will be known from QCD at the physical light-quark masses within the next
several years (if computational resources devoted to these calculations
continue to increase as they have during the last decade).  
These results will provide crucial input into the calculations of the structure and
interactions of light nuclei.  
The field of nuclear physics will be somewhat revolutionized by the deployment
of exa-scale computing resources as
they
will provide  predictive capabilities
that allow for the assignment of reliable uncertainty estimates in
observables that cannot be explored experimentally.
Further, they will enable the systematic exploration of
fundamental aspects of nature that are manifested in
the structure and interactions of nuclei.

\vskip 0.3in
{\it 
I would like to thank Amand Faessler and Jochen Wambach for inviting me to
participate in a most stimulating and enjoyable school.
I would also like to thank all of the members of the NPLQCD collaboration for
much of the progress described in this lecture.
}

\end{document}